\documentclass[twocolumn,tighten,usenames,dvipsnames,twocolappendix]{aastex63}
\usepackage[normalem]{ulem}
\usepackage{verbatim}
\usepackage{amssymb}
\usepackage{amsmath}
\usepackage{xspace}
\usepackage{graphicx}

\usepackage{amsmath,mathtools,mathrsfs}
\usepackage{hyperref}
\usepackage{afterpage}
\usepackage{booktabs}

\usepackage[encapsulated]{CJK}
\usepackage{ucs}
\usepackage[utf8x]{inputenc}

\newcommand{\abs}[1]{\left|#1\right|}

\def\G{{\rm G}} 

\def\muas{\mu{\rm as}} 
\def\Gl{{\rm G}\lambda} 

\def\ehtim{\texttt{eht-imaging}\xspace}

\def\erf{{\rm erf}}
\def\u{\textbf{u}}
\def\b{\textbf{b}}
\def\x{\textbf{x}}
\def\y{\textbf{y}}
\def\z{\textbf{z}}
\def\k{\textbf{k}}
\def\K{\textbf{K}}
\def\q{\textbf{q}}
\def\r{\textbf{r}}
\def\obs{{\rm obs}}
\def\intr{{\rm int}}
\def\Dp{D_\phi}
\def\H{\textbf{H}}

\def\as{\alpha_{\rm sc}}

\shorttitle{Probing Turbulence with EHT Polarimetry}
\shortauthors{Ni et al.}

\begin{document}

\title{
  Probing Accretion Turbulence in the Galactic Center with EHT Polarimetry
}

\correspondingauthor{Chunchong Ni}
\email{chunchong.ni@uwaterloo.ca}

\author[0000-0003-1361-5699]{Chunchong Ni}
\affiliation{ Department of Physics and Astronomy, University of Waterloo, 200 University Avenue West, Waterloo, ON, N2L 3G1, Canada}
\affiliation{ Waterloo Centre for Astrophysics, University of Waterloo, Waterloo, ON N2L 3G1 Canada}
\affiliation{ Perimeter Institute for Theoretical Physics, 31 Caroline Street North, Waterloo, ON, N2L 2Y5, Canada}

\author[0000-0002-3351-760X]{Avery E. Broderick}
\affiliation{ Perimeter Institute for Theoretical Physics, 31 Caroline Street North, Waterloo, ON, N2L 2Y5, Canada}
\affiliation{ Department of Physics and Astronomy, University of Waterloo, 200 University Avenue West, Waterloo, ON, N2L 3G1, Canada}
\affiliation{ Waterloo Centre for Astrophysics, University of Waterloo, Waterloo, ON N2L 3G1 Canada}

\author[0000-0003-2492-1966]{Roman Gold}
\affiliation{CP3 origins $|$ University of Southern Denmark (SDU) Campusvej 55, Odense, Denmark}

\begin{abstract}
Magnetic fields grown by instabilities driven by differential rotation are believed to be essential to accretion onto black holes.  These instabilities saturate in a turbulent state, and therefore the spatial and temporal variability in the horizon-resolving images of Sagittarius A* (Sgr A) will be able to empirically assess this critical aspect of accretion theory.  However, interstellar scattering blurs high-frequency radio images from the Galactic center and introduces spurious small-scale structures, complicating the interpretation of spatial fluctuations in the image.  We explore the impact of interstellar scattering on the polarized images of Sgr A*, and demonstrate that for credible physical parameters the intervening scattering is non-birefringent.  Therefore, we construct a scattering mitigation scheme that exploits horizon-resolving polarized mm/sub-mm VLBI observations to generate statistical measures of the intrinsic spatial fluctuations, and therefore of the underlying accretion-flow turbulence.  An optimal polarization basis is identified, corresponding to measurements of the fluctuations in magnetic field orientation in three dimensions.  We validate our mitigation scheme using simulated data sets, and find that current and future ground-based experiments will readily be able to accurately measure the image-fluctuation power spectrum.
\end{abstract}



\section{Introduction} \label{sec:I}

Black holes have been implicated as the engines of active galactic nuclei (AGN) and X-ray binaries. Within these objects, both their extreme luminosities and growth rate are presumably due to the interaction with the accretion of nearby matter. This occurs via accretion disks, through which material orbits, cools and falls inward toward the central object. Accretion flows are generic features in astronomical systems, from the formation of planets to the powering of AGN, and thus understanding the processes by which they operate informs astrophysics broadly.

The supermassive black hole at the Galactic Center, Sagittarius A* (Sgr A*), offers us a laboratory in which to study accretion flows in detail.  Located 8\,kpc from the Earth, with a mass of $4.3\times 10^6~M_\odot$ \citep{Ghez2016, Gillessen2009, Gravity2018}, Sgr A* has now been resolved on event horizon scales with the Event Horizon Telescope \citep[EHT][]{PaperI,PaperII,PaperIII,PaperIV,PaperV,PaperVI}.  These observations present an unprecedented opportunity to probe the nature and characteristics of the hot plasma orbiting Sgr A* under the extreme conditions near the horizon.

The EHT is a global array of millimeter and sub-millimeter telescopes that achieves resolutions of $20\,\muas$ at a wavelength of 1.3\,mm (230\,GHz) via very-long baseline interferometry (VLBI).  This resolution is sufficient to resolve the event horizons of Sgr A* and M87*, silhouetted against the emission from the surrounding hot plasma.  In comparison, the typical angular size of the shadow for Sgr A* and M87 is around $50 \muas$. Therefore, it has now become possible to probe accretion physics on scales comparable to those relevant for MHD turbulence. 


Four observing campaigns have been completed by the EHT, in April 2017, 2018, 2021 and 2022, and the first M87 and Sgr A* EHT results have been published \citep{2019ApJ...875L...1E,2019ApJ...875L...2E,2019ApJ...875L...3E,2019ApJ...875L...4E,2019ApJ...875L...5E,2019ApJ...875L...6E, PaperI, PaperII, PaperIII, PaperIV, PaperV, PaperVI}.  The full complement of Stokes parameters were measured at $230~{\rm GHz}$.  Future development of the array will include the ability to observe at $345~{\rm GHz}$. 


Sgr A* presents a natural target for studies of the role played by MHD turbulence in black hole accretion because of its short timescale and lack of an obvious relativistic jet.  However, interpreting the small-scale brightness fluctuations, presumably associated with MHD turbulence within the accretion flow, is complicated by the interstellar scattering observed toward the Galactic center \citep{Lo1998, Frail1994,Lazio&Cordes1998}.  This scattering is believed to be a result of variations in the electron density along the line of sight \citep{GoldreichSridhar1995, Lazio&Cordes1998b, Cordes&Lazio2001, Rickett1990}.  Typically, the origin of the scattering is abstracted to a thin scattering screen, for which detailed models exist \citep{Psaltis2018, Johnson2018, Issaoun2021, Cho2021}.  For Sgr A*, two aspects of the scattering are of interest, corresponding to different regimes: diffractive and refractive scattering \citep{Narayan1992, JohnsonGwinn2015}.



The diffractive scattering is the consequence of the combined effect of small-scale fluctuations in the interstellar electron density, whose impact is to blur the image with a nearly-Gaussian kernel \citep{JohnsonNarayan2016,Issaoun2021}. This angular broadening is formally reversible, i.e., images of Sgr A* may be effectively ``deblurred'' by applying the appropriate multiplicative correction in the Fourier (visibility) domain \citep{Fish2014, Lu2018, Johnson2015}.  

The impact of refractive scattering is more subtle.  Associated with the large-scale fluctuations of the interstellar electron density, refraction induces coherent and variable substructures in the image \citep{Johnson2018}.  These additional variations in the image are extrinsic to the source, and indicative of the interstellar scattering screen. Unlike diffractive scattering, it is not formally invertible, and may not be simply removed during image generation.  In principle, it may be modeled, leveraging the modestly different timescales between the refractively induced extrinsic substructure and the intrinsic brightness fluctuations induced by MHD turbulence.



In this paper, we demonstrate that the action of the scattering screen is expected to be independent of polarization.  Based on this, we develop a new scattering mitigation scheme that exploits this non-birefringence of the scattering screen.  We demonstrate that for the angular scales accessible to the EHT and ngEHT, it is possible to effectively eliminate the impact of interstellar scattering on the estimators of the intrinsic structural polarimetric fluctuations, and thus probe MHD turbulence instrinsic to the near-horizon emission region directly. While a full spatiotemporal characterization of the turbulence is highly desirable \citep[see, e.g.,][]{Georgiev_2022}, we focus on mitigating the spatial distortions resulting from scattering here, leaving the construction of the temporal component of the power spectrum for future work.


In \autoref{sec:theory} we review scattering in the thin-screen approximation, assess the impact on polarized emission, and demonstrate that scattering can be implemented as a nonbirefringent, tensor convolution that may be inverted.
In \autoref{sec:toy}, we construct toy models that mimic the gross properties of Sgr A*, and test the feasibility of scattering mitigation. In \autoref{sec:GRMHD}, we apply our scheme to a representative simulation from the existing EHT general relativistic magnetohydrodynamic (GRMHD) simulation library, and confirm we are able to extract intrinsic information about the structural variability in spite of the intervening scattering.  Finally, we conclude in \autoref{sec:C}.

\section{Scattering and Observation of Polarized Light} \label{sec:theory}


We begin with a summary of the action of an intervening scattering screen upon the emission from a compact source observed via a local interferometer. This is appropriate, e.g., for observations of Sgr A* by the EHT and ngEHT.  We will follow presentation of \citet{JohnsonGwinn2015} where possible and refer the reader there for a detailed description.

\subsection{Scattering and the Visibility Function}\label{sec:vis_def}

The primary observable quantity in interferometric radio observations, like those made by the EHT and ngEHT, is the ``visibility'', $V(\b)$, constructed by cross-correlating signals at antennae separated by a projected baseline $\b$.  This quantity is directly given by the Fourier transform of the intensity map, i.e., 
\begin{equation}
V(\b) = \int d^2\x e^{2\pi i \b\cdot\x / \lambda } I(\x),
\end{equation}
where $I(\x)$ is the intensity map projected at the source distance, $\x$ is an angular location on the sky, and $\lambda$ is the observing wavelength \citep[see, e.g.,][]{TMS}.  As a consequence, the $V(\b)$ encodes the degree of source structure on an angular scale of $\lambda/|\b|$, oriented along the direction of $\b$.

\begin{figure}
\begin{center}
\includegraphics[width=\columnwidth]{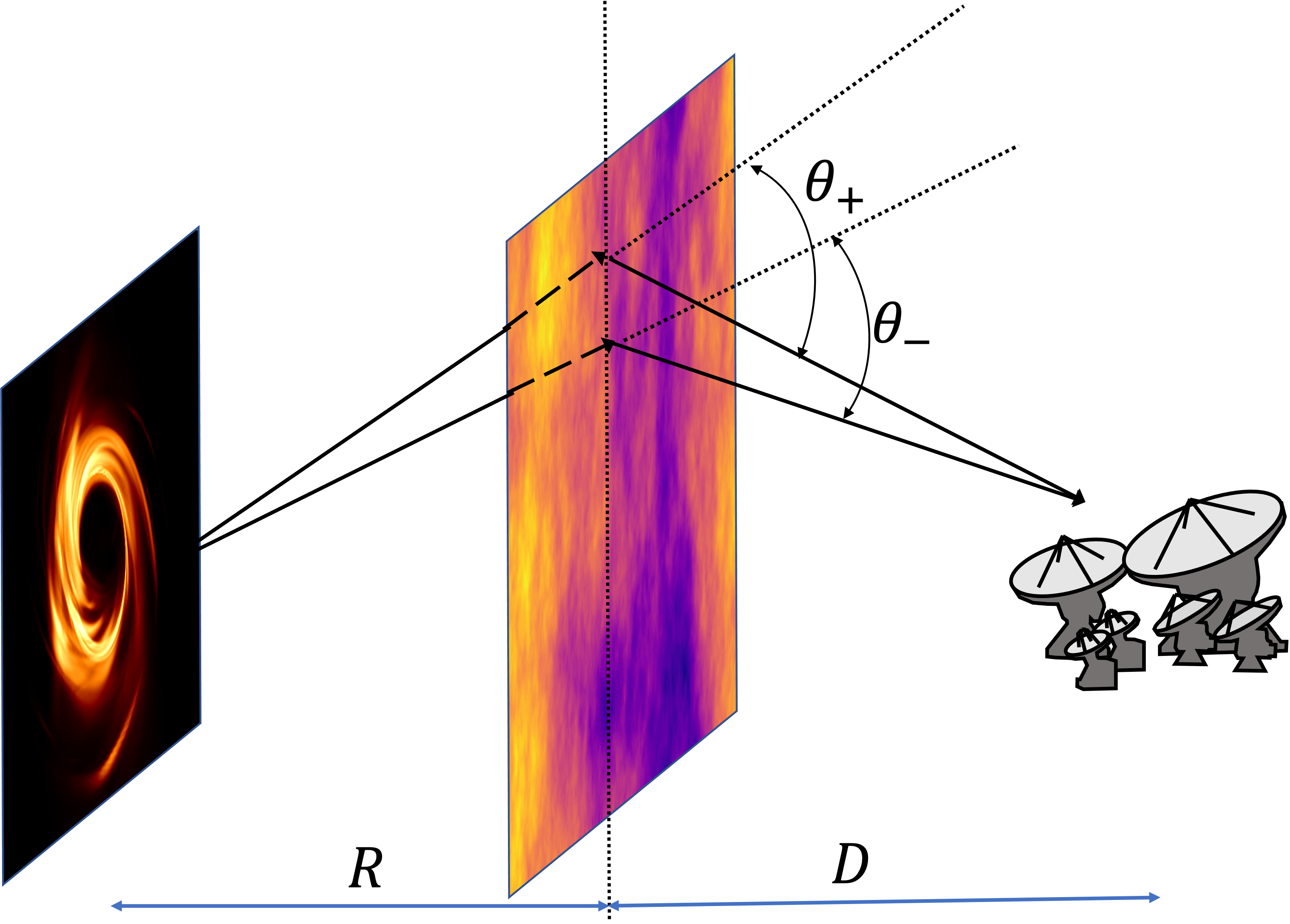}
\end{center}
\caption{Schematic illustration of the interstellar scattering at a thin screen that imparts random phase fluctuations. The paths of two rays, differing in polarization, are shown, indicating the slight difference in the refractive deflection angles $\theta_\pm$.  The source-screen ($R$) and screen-observer ($D$) distances are indicated.}
\label{fig:illustration} 
\end{figure}

Scattering is frequently modeled in the thin-screen approximation \citep{Bower}.  The physical picture is presented in \autoref{fig:illustration}, which shows the relative position of the emitting source, an intervening thin screen, and the observer on Earth.  Thick scattering screens include additional complication, and may be required toward Sgr A* (for example, see \cite{Ue-Li}).  Nevertheless, in many cases, these extended scattering regions may be abstracted to a sequence of thin screens.  Thus, we will focus on the latter.

The impact of scattering in the thin-screen limit, is to impart a random phase shift at the screen, $\phi(\x)$.  That is, the observed visibilities are
\begin{equation}
\begin{aligned}
V_{\rm obs}(\b) &=\frac{1}{4\pi^2 r_F^4}\int d^2\x_1 d^2\x_2\\
&\quad\times e^{i\left[\left(\x_1^2-\x_2^2\right)+\b/\left(1+M\right)\left(\b_1 +\x_2\right)\right]/\left(2r_F^{2}\right)}\\
&\quad\times 
e^{i\left[\phi\left(\x_1\right)-\phi\left(\x_2\right) \right]} V_{\rm int}\left[\left(1+M\right) \left(\x_2-\x_1\right)\right] . \label{visibility}
\end{aligned}
\end{equation}
where the Fresnel radius,
\begin{equation}
r_F = \sqrt{\frac{DR}{D+R} \frac{\lambda}{2\pi}}
\end{equation}
is the characteristic radius at the observer on which the spherical nature of the approaching radio wave become important, and provides a useful scale for scattering phenomena.  In \autoref{visibility}, we have introduced $V_{\rm obs}(\b)$ for the visibility that is observed after scattering and $V_{\rm int}(\b)$ for the visibility that would have been observed in the absence of scattering.  It is, fundamentally, $V_{\rm int}(\b)$ that is of interest to studies of the compact astronomical sources.

In the ensemble average regime, obtained after averaging 
$V_{\rm obs}(\b)$ over many realizations of the scattering screen, the observed visibility is given by
\begin{equation}
    \left< V_{\rm obs}(\b) \right>_{\rm ea}
    = 
    e^{-D_\phi(\b)} \left< V_{\rm int}[(1+M)\b] \right>_{\rm ea}
    \label{eq:diff_scatt}
\end{equation}
where $\langle\,\dots\rangle_{\rm ea}$ denotes ensemble averaging \citep{JohnsonGwinn2015}.  The $D_\phi(\b)$ is the structure function of the phase fluctuations on the scattering screen, defined in the normal way:
\begin{equation}
    D_\phi(\b) = \left< [ \phi(\x+\b)-\phi(\x) ]^2 \right>_{ea}
\end{equation}
where for a statistically isotropic screen, like that we assume here, the absolute position $\x$ does not matter.  \autoref{eq:diff_scatt} is the well-known diffractive limit, in which scattering imparts only a multiplicative correction to the appropriately averaged intrinsic visibilities, and in which images may be deblurred in the normal sense \citep{JohnsonGwinn2015}.

Analogous observable quantities can be constructed for polarized emission, and we do so for the Stokes maps, $S(\x)=[I(\x),Q(\x),U(\x), V(\x)]$, where here Stokes $V$ refers to the excess right-handed circular polarization, not the visibility\footnote{Henceforth, we will use superscripts to indicate Stokes parameters to avoid confusion.}.  In general, the phase shifts may depend on the particular polarization under consideration.  In practice, for credible values of the magnetic field strength in the interstellar medium, the scattering screen is non-birefringent.

\subsection{(Non-)Birefringence of Scattering in the ISM} \label{subsec:non-Bire}

The degree to which we may assume that an intervening scattering screen is non-birefringent depends on the magnitudes of two closely related quantities, the angular deflections experienced by radio waves passing through the screen (refraction), and the phase shifts imparted on those radio waves during their passage (dispersion).  Here, we show that for models of interstellar scattering in which both are due to turbulent fluctuations in the magnetized interstellar plasma, both are sufficiently small that we may treat scattering as independent of the polarization of the radio waves under consideration.  We will begin by analyzing the properties of the scattering in the unmagnetized limit, and thus produce estimates for the unpolarized case, followed by an analysis of a weakly magnetized screen to estimate the disparate impact on different radio wave polarizations. Before starting, we define the following dimensional variables, X and Y, following the conventions in plasma physics,
\begin{equation}
    X \equiv \omega_P^2/\omega^2,
\end{equation}
\begin{equation}
    Y \equiv {\rm sgn}(\textbf{B}\cdot\z) \omega_B/\omega,
\end{equation}
where $\omega_P = \left(4 \pi n_e e^2/m_e\right)^{1/2}$ is the plasma frequency, $n_e$ is the free electron density, $e$ the electron charge, and $m_e$ the electron mass, $\omega_B = eB/m_e c$ is the cyclotron frequency, and $\z$ is the line-of-sight direction. X and Y are proportional to the plasma density and the magnetic field strength at given frequency, $\omega$.

Detailed computations for screens composed of turbulent magnetized plasmas reach similar conclusions and are presented in \autoref{appendix:DvP}.

\subsubsection{Refraction in an Unmagnetized Screen}
In the absence of a magnetic field, the long-wavelength dispersion relation for electromagnetic waves in an electron-ion plasma is,
\begin{equation} \label{dispersionwomagnetic}
    \omega^2 = k^2 c^2 + \omega_P^2,
\end{equation}
 The associated equations of motion for radio wave are then given by Hamilton's equations obtained by setting $H(\x,\k)=\hbar\omega(\x,\k)$, 
\begin{equation}
    \dot{\textbf{k}} = -\vec{\nabla}\omega\left(\textbf{k},\x\right) = -\frac{\omega_P^2}{2\omega}\vec{\nabla}\ln{n_e},
\end{equation}
where $\textbf{k}$ is the wavevector and related to the photon momentum by $\textbf{p}=\hbar\textbf{k}$. This equation of motion describes how electromagnetic waves refract when travelling through the scattering screen. In the weak-deflection limit, the perpendicular momentum gained after propagating through the scattering screen is,
\begin{equation}
    k_\textbf{r} = \int dz \frac{dk_\textbf{r}}{dz} = -\int dz\frac{\omega_P^2}{2\omega c}\nabla_\textbf{r}\ln{n_e},
\end{equation}
where we have made use of the approximation that the line-of-site velocity of the wave is $c$. Thus, the deflection angle is approximately the ratio between $k_\textbf{r}$ and $k_z$,
\begin{equation}\label{theta0}
    \theta_0 = -\int dz  \frac{\omega_P^2}{2\omega^2}\nabla_\textbf{r}\ln{n_e}
    =
    -\frac{1}{2}\int dz \nabla_r X,
\end{equation}
where we used that $k_z\approx\omega/c$.

The typical size of scattered compact sources, broadened by the diffractive scattering, places a constraint on the magnitude of $\theta_0 \approx \theta_{\rm diff}$, and therefore, the line-of-sight integrated transverse gradients of the fluctuating electron density within the screen.  It is against this value that we will normalize the impact of non-zero magnetic fields, and thus the degree of birefringence for a weakly magnetized scattering screen.

\subsubsection{Refraction in a Weakly Magnetized Screen}
In the presence of a weak magnetic field, the dispersion relation of electromagnetic waves travelling through a plasma is slightly modified, becoming
\begin{equation}\label{DispersionwMagnetic}
    \omega^2 = k^2 c^2 + \omega_P^2 \left(1 \pm \frac{\omega_B}{\omega} \right),
\end{equation}
where there are now two propagating modes, ordinary and extraordinary, signified by the $\pm$ sign. In the quasi-transverse limit (propagation along the magnetic field), the two polarization modes are nearly circular.

We again obtain the equations of motion from Hamilton's equations, though in this instance they differ for the two polarization modes,
\begin{equation}
    \dot{\textbf{k}_\pm} = -\vec{\nabla}\omega\left(\textbf{k},\textbf{x}\right) = -\frac{1}{2\omega}\vec{\nabla}\left(\omega_P^2 \pm \frac{\omega_B}{\omega}\omega_P^2 \right).
\end{equation}
The deflection angle after integration through the screen is, therefore,
\begin{equation}\label{t}
    \theta_\pm = -\frac{1}{2}\int dz X\left(1\pm Y\right)\nabla_\textbf{r}\ln{\left[n_e\left(1\pm Y\right)\right]},
\end{equation}
where we have defined $Y$ previously.

The mean deflection of the two modes is just that associated with the unmagnetized plasma
\begin{equation}
\bar{\theta} = \frac{\theta_+ + \theta_-}{2} = \theta_0.
\end{equation}
The differential deflection, and thus the disparity in the impact of the weakly magnetized scattering screen on the two polarizations, is given by
\begin{equation}\label{dtheta}
    \delta \theta = \theta_+ - \theta_- = -\int dz\nabla_\textbf{r} \left(XY\right).
\end{equation}
Upon averaging over a random magnetic field orientation, and thus sign of $Y$, this would vanish.  However, the variance of $\delta\theta$, and thus its typical value, does not vanish,
\begin{equation}
\langle \delta\theta^2 \rangle
\approx
\langle\bar{\theta}^2\rangle \langle Y^2 \rangle>
\approx
\theta_{\rm diff}^2 \sigma_Y^2,
\end{equation}
where $\sigma_Y^2\sim Y^2$ is the variance in $Y$, associated with the magnitude of the magnetic field fluctuations.  Thus, typically, the difference in the deflection of the two polarization modes, is reduced by a factor of $Y^2$, which at 1.3~mm is small for credible interstellar magnetic field strengths ($\lesssim 1$~mG)

\subsubsection{Phase Shift Induced by a Weakly Magnetized Screen}
The typical phase fluctuations imparted by the scattering screen are related to, but distinct from, the refraction.  Given the dispersion relation in \autoref{DispersionwMagnetic}, the phase shift for the transverse electromagnetic modes grow as,
\begin{equation}
    \phi_\pm =\int kdz \approx \frac{\omega}{c} \int dz \left(1-\frac{1}{2}X \pm \frac{1}{2}XY \right),
\end{equation}
and thus the difference between the phases of the the polarization modes is,
\begin{equation}
    \delta \phi = \frac{\omega}{c} \int dz XY. 
\end{equation}
This expression suggests that the differentially accumulated phase and typical differential deflection angle differs by a factor of $L/\lambda$, where $L$ is the typical correlation length within the plasma.  This suspicion is born out by detailed calculations for a variety of magnetic field and electron density fluctuation spectra presented in \autoref{appendix:DvP}. Similar results are presented in the past literature (e.g. see \cite{MacqurtMelrose2000}).

In particular, for power-law electron fluctuation spectra, we show that $L$ may be associated with the inner scale, $r_{\rm in}$, in \cite{Psaltis2018} and \cite{Johnson2018}.  Therefore, the typical differential deflections may be related to the typical differential phase fluctuations induced by a thin scattering screen by,
\begin{equation}
    \langle \delta \phi^2 \rangle \approx \frac{r_{\rm in}^2}{\lambda^2}\langle \delta \theta^2 \rangle.
\end{equation}
Given a typical value for $\theta_{\rm diff}\approx10~\mu{\rm as}$, an interstellar magnetic field of $1~\mu{\rm G}$, and an inner scale within the Galactic center scattering screen of $r_{\rm in}\approx 800~{\rm km}$ \citep{Johnson2018,Issaoun2021}, we estimate that the root-mean-square phase difference between different polarization modes at 1.3~mm is of order
\begin{equation}
    \sqrt{\langle \delta \phi^2 \rangle} \approx 10^{-12} \left(\frac{B}{1 \rm{\mu G}}\right) \left(\frac{r_{\rm in}}{800~{\rm km}}\right)
    \,{\rm rad}.
\end{equation}
Note that the wavelength dependence of $\sqrt{\langle\delta \phi^2\rangle}$ is now nominally dropped. But the implicit dependence of the wavelength is within $\theta_{\rm diff}$, which is the typical value at $230 \rm GHz$. As a result, for the purposes of the EHT, we may safely assume that the Galactic center scattering screen is non-birefringent.

\subsection{Characterizing Turbulent Substructure in Images}

In this paper we are primarily interested in the measurement and characterization of the statistical properties of small-scale fluctuations in the underlying image of Sgr A*, presumably arising due to turbulent structures in the accretion and/or jet launching region. Typically, these are expressed in terms of power spectra, which measure the degree of fluctuations on each spatial scale.

As the Fourier transform of the sky brightness map, the $V(\b)$ already directly contain a measure of the degree of structure on various spatial scales.  Therefore, it is natural to construct statistical measures of the variability on different spatial scales from the $V(\b)$.  For reasons that will become clear in following sections, we choose to do this with the full visibilities, and characterize the spatial power spectrum of image fluctuations via,
\begin{equation}
    P^S(\b) = \left< |V^S(\b)|^2 \right>_{\rm turb},
    \label{eq:Pdef}
\end{equation}
where the superscript $S$ indicates which Stokes parameter is used to construct the visibilities, and $\langle\,\dots\rangle_{\rm turb}$ indicates averages over timescales long in comparison to the turbulent timescales in the source on the spatial scales of interest (typically many hours or longer).  We will presume henceforth that all time averages will include both turbulent averages and ensemble averages, i.e., independent averages over realizations of the stochastic source structure and the intervening scattering screen, dropping the specifier in what follows.

Note that this does not subtract the mean $V(\b)$, and therefore contains contributions from the variable and static components of the image.  Nevertheless, we will find that this is the more convenient power spectrum for scattering mitigation. It is also defined consistently with the power spectral densities in \cite{Georgiev_2022}.


\subsubsection{Impact of Scattering on the Power Spectrum}

After scattering via a thin screen, as described in \autoref{sec:vis_def}, the power spectrum is modified.  The resulting expression may be found in Eq.\ 32 of \citet{JohnsonGwinn2015}, which reads
\begin{multline}
P^S_\obs(\b) = 
e^{-\Dp[\b/(1+M)]} P^S_\intr(-\b)\\
-
\frac{1}{(2\pi r_F^2)^2}
\frac{1}{2 r_F^4}
\int d^2\y\,
\left[ \y\cdot\left(\y+\frac{\b}{1+M}\right)\right]^2\\
\times
\tilde{D}_\phi\left( \y + \frac{\b}{1+M}\right) 
e^{-\Dp(\y)} P^S_\intr[(1+M)\y],
\label{eq:PS_scatt}
\end{multline}
where 
\begin{equation}
\tilde{D}_\phi(\y)
=
\int d^2\x\, e^{i \y\cdot\x/r_F^2} D_\phi(\y)
\end{equation}
is the Fourier transform of the structure function.  

Hidden within \autoref{eq:PS_scatt} is a convolution that, like for diffractive scattering, expresses the impact of refractive scattering as a linear operator, defined by the tensor convolution kernel,
\begin{equation}
\K(\y) = - \frac{\y \tilde{D}_\phi[\y/(1+M)] \y^T}{8\pi^2 r_F^8(1+M)^6}.
\label{eq:Kdef}
\end{equation}
Expressing \autoref{eq:PS_scatt} in terms of $\K(\y)$, we obtain,
\begin{multline}
P^S_\obs(\b) = 
e^{-\Dp[\b/(1+M)]} P^S_\intr(-\b)\\
\int d^2\y\,
\sum_{ij} \K\left(\y+\b\right)_{ij} \left[ \y e^{-\Dp[\y/(1+M)]}P^S_\intr(\y) \y^T \right]_{ji} ,
\label{eq:scattconv}
\end{multline}
where both indices of $\K(\y)$ are summed over. \autoref{eq:scattconv} is equivalent to Eq.\ 16 of \citet{JohnsonNarayan2016} after identifying $\tilde{D}_\phi$ with their $Q$, up to appropriate scalings.  Within \autoref{eq:scattconv}, the impact of diffractive and refractive scattering are clearly delineated by the first and second terms, respectively.

\subsubsection{Characteristic scales for $\K(\y)$} \label{sec:exp_Dphi}

The typical scales for $\K(\y)$ may be inferred from its definition and the approximate limiting expressions for $\Dp(\x)$, and thus $\tilde{D}_\phi(\y)$.  
On very small scales, $\Dp(\x)$ is generically quadratic, smoothly vanishing at $\x=0$.  Assuming isotropy of the phase screen, on very large scales, $\Dp(\x)$ is a power law fixed by the nature of the turbulence within the ISM that gives rise to the scattering screen.  Therefore, following \citet{JohnsonGwinn2015}, we express $\Dp(\x)$ in terms of these two regimes, separated by a spatial scale within the scattering screen, $r_0$, which we will assume is much smaller than the scale at which the ISM turbulence is damped, the ``inner scale'', $r_{\rm in}$:
\begin{equation} \label{StructureFunction}
D_\phi(\x)
=
\begin{cases}
\displaystyle
\left(\frac{|\x|}{r_0}\right)^2 & |\x| \ll r_{\rm in}\\
\displaystyle
\frac{2}{\as} \left(\frac{r_{\rm in}}{r_0}\right)^{2-\as}
\left(\frac{|\x|}{r_0}\right)^{\as}
& |\x| \gg r_{\rm in}.
\end{cases}
\end{equation}

Typically, the longest baselines accessible to the EHT array are well into the power-law regime, assuming the inner scale of $800 \rm{km}$.

In this case, the above expressions simplify to
\begin{equation}
D_\phi(\x)
\approx
\left(\frac{|\x|}{r_{\rm diff}}\right)^{\as},
\label{eq:Dphi}
\end{equation}
where
\begin{equation}
r_{\rm diff} = r_{\rm in} \left( \frac{\as}{2} \right)^{1/\as} \left( \frac{r_0}{r_{\rm in}} \right)^{2/\as}.
\end{equation}
The corresponding $\tilde{D}_\phi(\y)$ is given in Eq.\ 34 of \citet{JohnsonGwinn2015},
\begin{multline}
\tilde{D}_\phi(\y)
=
2^{2+\as} \pi
\frac{\Gamma(1+\as/2)}{\Gamma(-\as/2)}
r_{\rm diff}^2
\left(
\frac{r_{\rm diff}}{r_F}
\frac{|\y|}{r_F}
\right)^{-(\as+2)}.
\label{eq:Dphitilde}
\end{multline}
The collection of constants in front of $r_{\rm diff}^2$ evaluate to -7.09 for $\as=1.38$, appropriate for Sgr A* \citep{Issaoun2021}.  
Inserting this into \autoref{eq:Kdef}, the refractive scattering kernel is approximately
\begin{multline}
\K(\y)
\approx
-\frac{2^{\as-1}}{\pi} \frac{\Gamma(1+\as/2)}{\Gamma(-\as/2)}
\frac{1}{r_F^4  (1+M)^{4-\as}}\\
\times
\left(
\frac{r_{\rm diff}}{r_F}
\right)^{-\as}
\frac{\y}{r_F}
\left(\frac{|\y|}{r_F}\right)^{-(\as+2)}
\frac{\y^T}{r_F}.
\label{eq:Keval}
\end{multline}

\subsubsection{Approximate Inversion of Refractive Scattering} \label{sec:first_order}

\autoref{eq:scattconv} provides $P^S_\obs(\b)$ in terms of a linear operation upon $P^S_\intr(\b)$.  Because it is the latter that is of particular interest here, we need to invert this relation, giving $P^S_\intr(\b)$ in terms of $P^S_\obs(\b)$.  How to do this is discussed in detail in \autoref{app:scatt_inverse}.  If the refractive term is small, this inversion can be constructed perturbatively, yielding to first order
\begin{multline}
P^S_\intr(\b)
=
e^{D_\phi[\b/(1+M)]} \bigg\{
P^S_\obs(\b)\\
-
\int d^2\y\, \sum_{ij} \K(\y+\b)_{ij} \left[\y P^S_\obs(\y)\y^T \right]_{ji}
\bigg\}.
\label{eq:Pinv_linear}
\end{multline}
The accuracy of this approximation is dependent on the magnitude of ${P^S_\obs(\b)}^{-1} \int d^2\y\, \sum_{ij} \K(\y+\b)_{ij} \left[\y P^S_\obs(\y)\y^T \right]_{ji}$, which must be small.  This ratio is approximately the fraction of the observed power due to refractive scattering, which is what we are explicitly expanding in.

It also indicates the origin of the scattering-mitigation strategy pursued by combining multiple polarization modes.  The idea is to minimize this ratio.  It is now clear how this may be done.  Very red or blue $P^S_\obs(\b)$ will distribute power from large or small scales, respectively, throughout $P^S_\intr(\b)$.  Therefore, suppressing strong variations with spatial frequency in $P^S_\obs(\b)$ is primary way in which the choice of polarization mode can impact the fidelity of the $P^S_\intr(\b)$ reconstruction.

At the same time, it explicitly identifies that this approximation cannot be satisfied at all baseline lengths.  At sufficiently long baselines, $P^S_\obs(\b)$ is exponentially suppressed by the diffractive scattering.  As a result, the power distributed from short baselines by the convolution term will inevitably dominate.  As a result, it is exponentially difficult to push toward longer baselines in the exponentially suppressed regime.  If we define $\b_{\rm diff}$ to be the shortest baseline for which $D_\phi[\b_{\rm diff}/(1+M)]\approx1$, then the perturbative expansion will be poorly justified for baseline lengths above $|\b_{\rm max}|$, defined by
\begin{multline}
\sum_{ij} \K(\b_{\rm diff}+\b_{\rm max})_{ij} \left[
\b_{\rm diff} P^S_\intr(\b_{\rm diff}) \b_{\rm diff}^T
\right]_{ji} |\b_{\rm diff}|^2\\
\approx
e^{-D_\phi[\b_{\rm max}/(1+M)]} P^S_\intr(\b_{\rm max}).
\end{multline}

If we assume that at long baselines the intrinsic spatial power spectrum has a power-law fall off, i.e., $P^S_\intr(\b)\sim|\b|^{-\alpha}$, then with $D_\phi(\b)$ and $\tilde{D}_\phi(\y)$ given by \autoref{eq:Dphi}, \autoref{eq:Dphitilde}, and \autoref{eq:Keval}
\begin{equation}
|\K(\b_{\rm max})|
\approx
\frac{K_0}{64 r_F^4} 
\left(\frac{b_{\rm diff}}{r_F}\right)^{-2\as}
\left(\frac{b_{\rm max}}{b_{\rm diff}}\right)^{-\as},
\end{equation}
where $K_0$ encapsulates the $\as$-dependent coefficient preceding the factor of $r_F^{-4}$ in \autoref{eq:Keval}.
Therefore, the condition that the refractive and diffractive contributions to $P^S_\obs$ are similar becomes,
\begin{equation}
\frac{K_0}{64}
\left(\frac{b_{\rm diff}}{r_F}\right)^{4-2\as}
\left(\frac{b_{\rm max}}{b_{\rm diff}}\right)^{\alpha-\as}
\approx
e^{-[b_{\rm max}/b_{\rm diff}(1+M)]^{\as}},
\end{equation}
and therefore, the maximum baseline length at which we may perturbatively invert \autoref{eq:scattconv} to obtain $P^S_\intr$ is
\begin{equation}
b_{\rm max} \approx (1+M) b_{\rm diff} \left[
\ln\left(\frac{64}{K_0}\right)
+(4-2\as) \ln\left(\frac{r_F}{b_{\rm diff}}\right)
\right]^{1/\as},
\label{eq:estbmax}
\end{equation}
where we have ignored a logarithmic term that scales as $(\alpha-\as)\ln(b_{\rm max}/b_{\rm diff})$.\footnote{Explicitlty including this term does not significantly change the approximate limit.}

For $\as=1.38$, $R=2~{\rm kpc}$ and $D=6~{\rm kpc}$, $r_F\approx10^5~{\rm km}$, and $b_{\rm min}\approx3~{\rm G}\lambda$ (corresponding to the long-axis of the diffractive scattering kernel at 1.3~mm), this gives that $b_{\rm max} \approx 3.4 b_{\rm diff}(1+M) \approx 13.6~{\rm G}\lambda$.

Note that because $b_{\rm diff}$ grows as $\lambda^{-2}$ and $r_F$ grows as $\lambda^{1/2}$, $b_{\rm max}$ grows more rapidly than the maximum baseline as $\lambda$ shrinks.  For example, if $\lambda$ decreases from 1.3~mm to 0.87~mm, $b_{\rm diff}$ grows by a factor of 2.2, while the remainder of the coefficient in \autoref{eq:estbmax} decreases to 3.1, and thus $b_{\rm max}$ increases to $32.8~{\rm G}\lambda$.  In comparison, the maximum Earth-bound baseline grows from $8.5~{\rm G}\lambda$ to $12.8~{\rm G}\lambda$.  Hence, at 0.87~mm, the linear approximation improves dramatically.

Henceforth, we will assume that we may utilize the first order inversion approximation in \autoref{eq:Pinv_linear} to recover $P^S_\intr(\b)$ from observations of $P^S_\obs(\b)$.

\subsection{Exploiting Polarization}
\label{subsubsec:ratio}

The dominant impact of scattering on $P^S_\intr(\b)$ is the reduction of power at long baselines due to diffractive scattering.  However, because the action of scattering is independent of the polarization of the observed radio wave, combinations of polarized power spectra may be constructed such that the diffractive suppression is cancelled identically.  That is, with \autoref{eq:Pinv_linear} applied to the power spectra measured for Stokes parameters $S$ and $S'$, we have
\begin{equation}
\frac{P^{S}_\intr(\b)}{P^{S'}_\intr(\b)} =
\frac{
  P^{S}_\obs(\b)-\int d^2\y\, \sum_{ij} \K(\y+\b)_{ij} [\y P^{S}_\obs(\y)\y^T]_{ji}
}{
  P^{S'}_\obs(\b)-\int d^2\y\, \sum_{ij} \K(\y+\b)_{ij} [\y P^{S'}_\obs(\y)\y^T]_{ji}
},
\label{eq:Pratio}
\end{equation}
which is impacted only by refractive scattering.  In this way, diffractive scattering can effectively be mitigated, even without an explicit model for the scattering process (i.e., a specific choice of $\Dp(\x)$).

The reconstruction of $P^S_\intr(\b)$ from $P^{S'}_\obs(\b)$ in \autoref{eq:Pinv_linear} improves dramatically when $P^S_\obs(\b)$ is small on short baselines, and therefore there is less refractive contamination at long baselines to the recovered intrinsic power spectrum.  This is reflected by the smaller contributions from the convolution term in \autoref{eq:Pinv_linear}.  When this may be neglected, \autoref{eq:Pratio} reduces to
\begin{equation}
\frac{P^{S}_\intr(\b)}{P^{S'}_\intr(\b)} \approx \frac{P^{S}_\obs(\b)}{P^{S'}_\obs(\b)}.
\label{eq:Pratio_approx}
\end{equation}
Given the measurement of the full Stokes maps, it is generally possible to construct specific polarization modes for which the assumptions underlying \autoref{eq:Pratio_approx} are satisfied. 


Because the $P^S_\obs(\b)$ are typically ``red'', it will suffice to select the polarization modes to preferentially suppress the short-baseline (large-scale) power.  By doing so, the contamination of the estimated $P^S_\intr(\b)$ at long baselines (small scales) from the $P^S_\obs(\b)$ at short baselines (large scales) can be nearly eliminated.  That is, 
by minimizing the large scale power in the $P^S_\intr(\b)$, we are able to minimize the magnitude of the ratio of ${P^S_\obs(\b)}^{-1} \int d^2\y\, \sum_{ij} \K(\y+\b)_{ij} \left[\y P^S_\obs(\y)\y^T \right]_{ji}$ in \autoref{eq:Pratio}, rendering \autoref{eq:Pratio_approx} an excellent approximation at sufficiently long baselines.

\begin{figure}
\begin{center}
\includegraphics[width=\columnwidth]{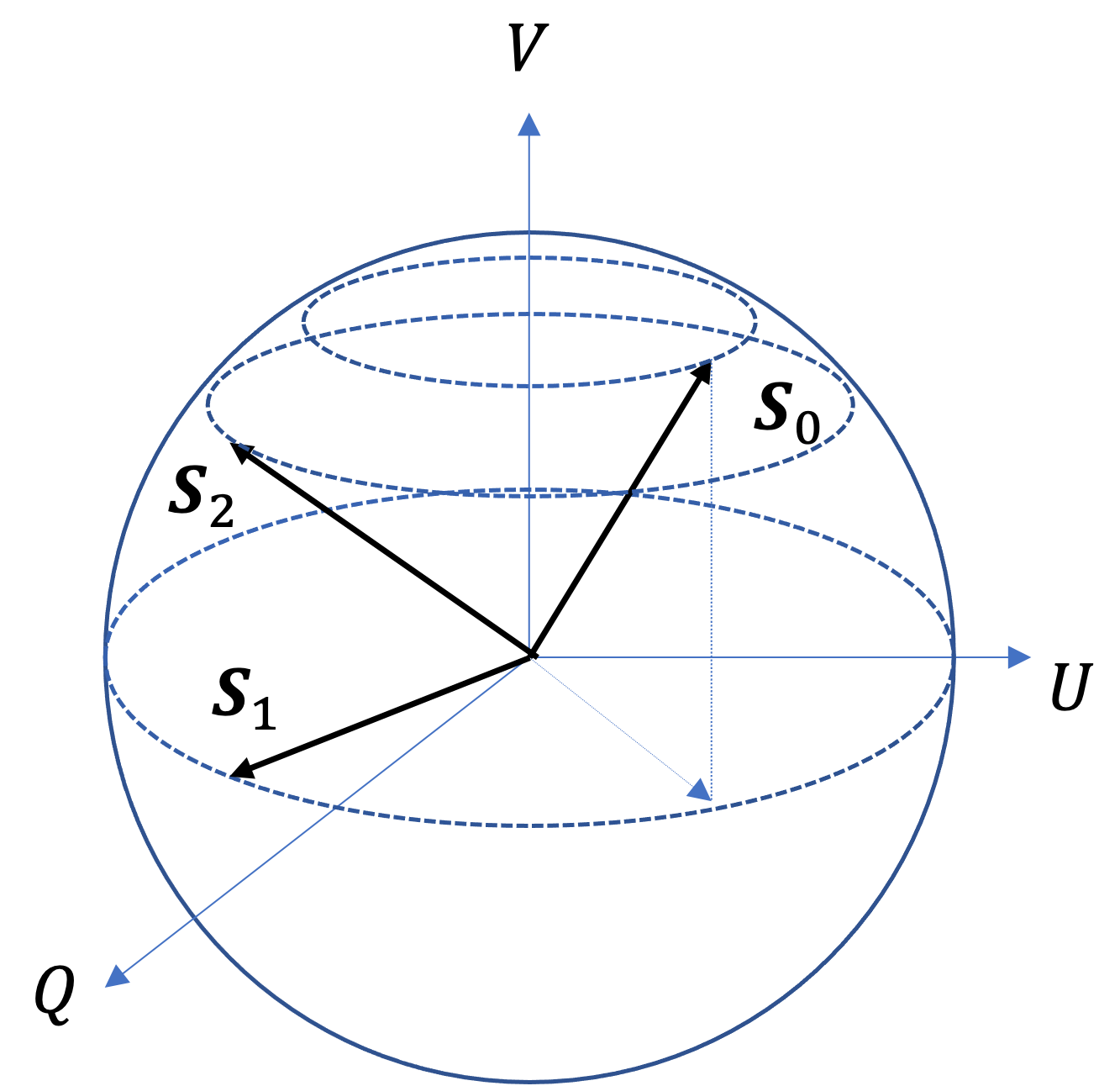}
\end{center}
\caption{Relative orientations of $\textbf{S}_0$, $\textbf{S}_1$ and $\textbf{S}_2$ are shown on the Poincare sphere.  $\textbf{S}_1$ is orthogonal to $\textbf{S}_0$ and is restricted to lie the $Q$-$U$ plane.  $\textbf{S}_2$ is orthogonal to both $\textbf{S}_0$ and $\textbf{S}_1$.  See the text for explicit definitions.
}
\label{fig:StokesSphere} 
\end{figure}

We begin by constructing the Stokes vector, $\textbf{S}_0$, associated with the time-averaged values of zero-baseline Stokes visibility maps $Q$, $U$, and $V$ (corresponding the source-integrated polarization).  This vector is shown in \autoref{fig:StokesSphere} after projecting it onto the Poincare sphere.  We construct two additional polarization modes, $\textbf{S}_1$ and $\textbf{S}_2$, chosen to be orthogonal to $\textbf{S}_0$.  As a direct result, $\langle \textbf{S}_1 \rangle = \langle \textbf{S}_2 \rangle = 0$ identically, and therefore $P^{S1}_\obs(0)$ and $P^{S2}_\obs(0)$ are generally small.

We choose to construct $\textbf{S}_1$ from $\langle V^Q \rangle(0)$ and $\langle V^U \rangle(0)$, thereby ensuring that it corresponds to a linearly polarized mode.  The resulting $\textbf{S}_1$, after enforcing orthogonality with $\textbf{S}_0$ is shown in \autoref{fig:StokesSphere}.  The second polarization is then unique defined up to a sign by the requirement that $\textbf{S}_2$ be orthogonal to both $\textbf{S}_1$ and $\textbf{S}_0$, as shown in \autoref{fig:StokesSphere}.  Generally, $\textbf{S}_2$ will be an elliptical polarization mode.  

Explicitly, in terms of $\textbf{S}_0 = \left( q_0, u_0, v_0 \right)$, $\textbf{S}_{1}$ and $\textbf{S}_{2}$ can be expressed as
\begin{equation}
\textbf{S}_{1} = \left( q_1, u_1, 0 \right)
~~\text{and}~~
\textbf{S}_{2} = \left( q_2, u_2, v_2 \right),
\end{equation}
where $q_1$, $u_1$, $q_2$, $u_2$ and $v_2$ are the projected components of the Stokes vector onto $Q$, $U$ and $V$ axes:
\begin{equation}
\begin{aligned}
    q_1 &= -u_0 / \sqrt{q_0^2+u_0^2}\\
    u_1 &= q_0 / \sqrt{q_0^2+u_0^2}\\
    q_2 &= -u_1 v_0 / \sqrt{q_0^2+u_0^2+v_0^2}\\
    u_2 &= q_1 v_0  / \sqrt{q_0^2+u_0^2+v_0^2}\\
    v_2 &= (q_1^2+u_1^2) / \sqrt{q_0^2+u_0^2+v_0^2}
\end{aligned}
\end{equation}


The power spectra for the two polarization modes can be written down as the linear combination of the power spectra associated with the polarization components $Q$, $U$ and $V$, with the coefficients $q_1$, $u_1$, $q_2$, $u_2$ and $v_2$, i.e.,
\begin{equation}
\begin{aligned}
    P^{\textbf{S}_1} &= q_1^2 P^Q + u_1^2 P^U + q_1 u_1 \langle V^Q V^{U*} \rangle + u_1 q_1 \langle V^U V^{Q*} \rangle \\
    P^{\textbf{S}_2} &= q_2^2 P^Q + u_2^2 P^U + v_2^2 P^V\\
    &+ q_2 u_2 \langle V^Q V^{U*} \rangle + u_2 q_2 \langle V^U V^{Q*} \rangle \\
    &+ q_2 v_2 \langle V^Q V^{V*} \rangle + v_2 q_2 \langle V^V V^{Q*} \rangle \\
    &+ u_2 v_2 \langle V^U V^{V*} \rangle + v_2 u_2 \langle V^V V^{U*} \rangle
\end{aligned}
\end{equation}

The benefits of these two constructed polarization modes are two-fold. First, the Stokes vectors of $\textbf{S}_1$ and $\textbf{S}_2$ are perpendicular to $S_0$, so they both have zero-mean identically, which ensures minimal impact from scattering. 
First, the absence of a mean value for $\textbf{S}_1$ and $\textbf{S}_2$ renders \autoref{eq:Pratio_approx} an excellent approximation of the relationship between the observed and intrinsic power spectra.
Second, they have a clear physical meaning.

Fluctuations in $\textbf{S}_1$ correspond to variations in the observed electric vector position angle (EVPA), i.e., the orientation of the linear component of the polarization.  For synchroton sources, as Sgr A* is believed to be, this maps the projected orientation of the net magnetic field as measured on different spatial scales.  Fluctuations in $\textbf{S}_2$ correspond to variations in the observed ellipticity, i.e., the degree of circular polarization relative to that of the linear polarization.  For synchroton emission from ion-electron plasmas, again, anticipated to be the case in Sgr A*, this directly maps to the angle between the magnetic field and the line of sight.  Therefore, these two polarization modes are intrinsically probing the stochastic variability in the three-dimensional magnetic field, integrated throughout the emission region.  

MHD turbulence is expected to generate large variations in the plasma density, magnetic field strength, and magnetic field orientation.  Thus, given global simulations of the emitting plasma, testable predictions for $P^{S1}_\intr(\b)$ and $P^{S2}_\intr(\b)$, and hence their ratio, may be generated.  That is, not only is $P^{S1}_\intr(\b)/P^{S2}_\intr(\b)$ technically easier to measure, but it is precisely the quantity that is expected to provide direct insight into the astrophysical processes within the source.

\subsection{Signatures of Temporal Variability}
In \autoref{eq:Pdef}, we intentionally did not construct the more common power spectrum of the fluctuations about the mean, $\mu^S(\b) \equiv \left< V^S(\b) \right>_{\rm turb}$.  This is because the impact of scattering on $P^S(\b)$ and $\mu^S(\b)$ are very different, with
\begin{equation}
    \mu_\obs^S(\b) = e^{-\Dp[\b/(1+M)]/2} \mu^S_\intr(\b)
\end{equation}
\citep{JohnsonNarayan2016}.
Nevertheless, observed and intrinsic estimates for ratios of the $\mu^S(\b)$ may also be constructed:
\begin{equation}
    \frac{\mu_\obs^S(\b)}{\mu_\obs^{S'}(\b)}
    =
    \frac{\mu_\intr^S(\b)}{\mu_\intr^{S'}(\b)}
\end{equation}

and their relationship to the corresponding ratios of $P^S(\b)$ provide evidence for intrinsic source variability.

If the intrinsic source is stationary, 
\begin{equation}
    P_\intr^S(\b) = |\mu_\intr^{S}|^2(\b),
\end{equation}
and the ratio of $P^S(\b)$ and $|\mu^{S}|^2(\b)$ are identical.  In contrast, when the intrinsic source is variable,
\begin{equation}
    P_\intr^S(\b) = |\mu_\intr^{S}|^2(\b) + \Sigma_S^2(\b)
\end{equation}
where $\Sigma_S^2(\b)$ is the variance due to temporal fluctuations in the source structure on spatial scales $\b$.  Thus, when the source is intrinsically variable, $P_\intr^S(\b)$ is strictly larger than $|\mu_\intr^{S}|^2(\b)$.  Unfortunately, in the absence of prior knowledge about the nature and magnitude of the variability in the two Stokes parameters, we do not know a priori if the ratios are larger, smaller or equal.  Nevertheless, if
\begin{equation}
    \frac{P^{S}_\obs(\b)}{P^{S'}_\obs(\b)}
    \ne
    \frac{|\mu^{S}_\obs|^2(\b)}{|\mu^{S'}_\obs|^2(\b)},
\end{equation}
at a statistically significant degree, then the source must be intrinsically temporally variable.  That is, the variability cannot simply be due to the impact of scattering.

\section{Validation with Simple Source Structure}
\label{sec:toy}
In this section, we present numerical experiments of scattering mitigation with simple source structures for which we have full control of the relevant power spectra.  The purpose of this section is to demonstrate:
\begin{enumerate}
    \item The ability to construct simple source structures with reasonable power spectra that are similar to the target source, Sgr A*,
    \item That the approximation in \autoref{eq:Pratio_approx} is well justified and successfully permits reconstruction of probes of the intrinsic variability.
\end{enumerate}

We begin with a description of how toy image models with different power spectra and polarization modes are constructed.  This is followed by a set of simulated observations in which scattering is incorporated using \ehtim \citep{EHTim_paper, EHTim_software}. Power spectra are constructed and \autoref{eq:Pratio} for various choices of $S$ and $S'$ are compared for the intrinsic (pre-scattered) and observed (post-scattered) images.  In all cases we assumed a wavelength of 1.3\,mm.


\subsection{Constructing Structured Intrinsic Image} \label{subsec:ToyModelIntrinsic}

The toy model is comprised of a Gaussian delta ring envelope and set of over-imposed fluctuations with a known power spectrum. 
In more detail:
\begin{enumerate}
    \item A mean background image is chosen, $G(\x)$.  For all experiments reported in this section, we adopt a Gaussian delta ring with radius of $25\,\muas$ and with width of $5\,\muas$
    \item A power spectrum for the fluctuations is chosen, i.e., 
    \begin{equation}
        \wp(\textbf{k};\sigma_P, \alpha)
        =
        \sigma_\wp^2 
        \left[ 1 + \left( \frac{\textbf{k}X}{2\pi}\right)^2 \right]^{\alpha/2},
    \end{equation} 
    for some normalization $\sigma_\wp$ and fluctuation spectral index $\alpha$, where $X$ is some maximum spatial scale.  Because both Sgr A* and GRMHD simulations exhibit fluctuations dominated by those on the largest spatial scales, we will assume that $\alpha<0$ generally.
    \item  A realization of fluctuations are constructed from a set of zero-mean, unit variance Gaussian random variables (GRVs).  That is, on a grid in the Fourier domain, at each $\textbf{k}$ we choose two GRVs, $\mathcal{N}_1$ and $\mathcal{N}_2$, from which the Fourier components of the fluctuation map are given by
    \begin{equation}
        F_{\textbf{k}}
        =
        \sqrt{\frac{\wp(\textbf{k})}{2}} \left( \mathcal{N}_1 + i\mathcal{N}_2\right),
    \end{equation}
    where we have suppressed the remaining arguments of the fluctuation power spectrum.  The spatial $f(\x)$ is constructed by the FFT of the $F_{\textbf{k}}$.
    \item In the image domain, the desired model for the total intensity is obtained via
    \begin{equation}
        I(\x) = G(\x)e^{f(\x)},
    \end{equation}
    where we exponentiate $f(x)$ to ensure positivity.
    \item Polarized images are generated in a similar fashion as described above, with two additional Gaussian random fields, $l_1(\x)$ and $l_2(\x)$ constructed similarly to $f(\x)$ but with independent $\alpha_{L1}$ and $\alpha_{L2}$, from which
    \begin{equation}
        L_1(\x) = l_1(\x) I(\x)
        ~~\text{and}~~
        L_2(\x) = l_2(\x) I(\x).
    \end{equation}
    These have zero mean by construction.
\end{enumerate}

The definition and the setup the toy model are controlled by five parameters, which encodes five aspects of the desired properties of the toy model: the amplitude of the fluctuations, $\sigma_\wp$, the spectral indexes of the power spectra for the total intensity, $\alpha$, and two polarization models $\alpha_{L1}$ and $\alpha_{L2}$, and a spatial scale on which power spectrum flattens, $X$, which we will set to $25\,\muas$.



Now, we have successfully generated the toy model, with both the total intensity and the polarized components, and the next step is to simulate observations.


\subsection{Simulated Ensemble of Observations}
\label{subsec:ToyModelObs}

Observations of Sgr A* are impacted by two additional effects: variability and the interstellar scattering we seek to mitigate.  To apply the scattering we make use of the Stochastic Optics package within \ehtim, which implements the scattering model described in \citet{Johnson2016} with the parameters measured in \citet{Johnson2018} and \citet{Issaoun2021}.  For all simulated Stokes map ($I$, $L_1$, and $L_2$), a single realization of the scattering screen is employed for each simulated instantaneous image.

Variability, both within the source and the scattering screen, is incorporated by producing a large collection of scattered images, each with a unique randomly constructed scattering screen and set of intrinsic fluctuations.  In this way, we generate a statistical ensemble of observed images.  Note that this procedure ignores potential temporal correlations within the image that will present themselves in the following section, when GRMHD models are considered.

Specifically, in this numerical experiment, we chose the following values for the parameters for this simple source structure model. The amplitude of the fluctuation of the Gaussian delta ring envelope, $\sigma_\wp$, is 100. The power index for the total intensity, $\alpha$, is $-4$. The power indices associated with the two polarization modes, $\alpha_{L1}$ and $\alpha_{L2}$, are $-4$ and $-2$, respectively. In conclusion, the simple source model has a red power spectrum, dominated by large-scale fluctuation. The associated two polarization modes also have independent red spectra, but with different power indices.

With this source model, we first calculate its intrinsic visibility, for the total intensity and the polarization modes, noted as $V^{I}_\intr(\u)$, $V^{L1}_\intr(\u)$ and $V^{L2}_\intr(\u)$. Second, we generate the scattered image using the scattering model implemented in $\ehtim$. Each snapshot of the source is scatted with an independent realization of the scattering screen. Third, we calculate the scattered visibility, noted as $V^{I}_\obs(\u)$, $V^{L1}_\obs(\u)$ and $V^{L2}_\obs(\u)$. Then, the intrinsic and scattered visiblities are averaged down over 100 realizations of the source and the scattering screen. Finally, we take the ratios among averaged scattered visibilities and averaged intrinsic visibilites, respectively.

For this first set of validation tests, we ignore measurement uncertainties, e.g., those associated with thermal fluctuations.  Thus, the primary source of uncertainty is sampling error associated with the finite number of simulated images in the ensemble.  In \autoref{sec:GRMHD}, we include more realistic assessments of array performance for EHT and ngEHT.



\subsection{Power Spectra Estimation Results}
\label{sec:simple_results}

\begin{figure*}
\begin{center}
\includegraphics[width=0.329\textwidth]{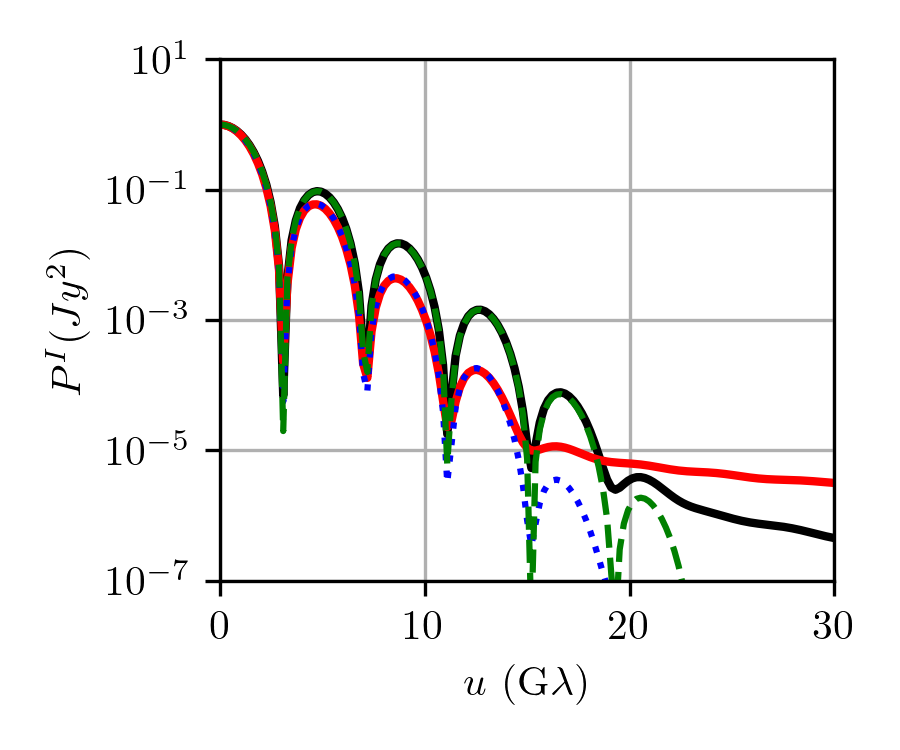}
\includegraphics[width=0.329\textwidth]{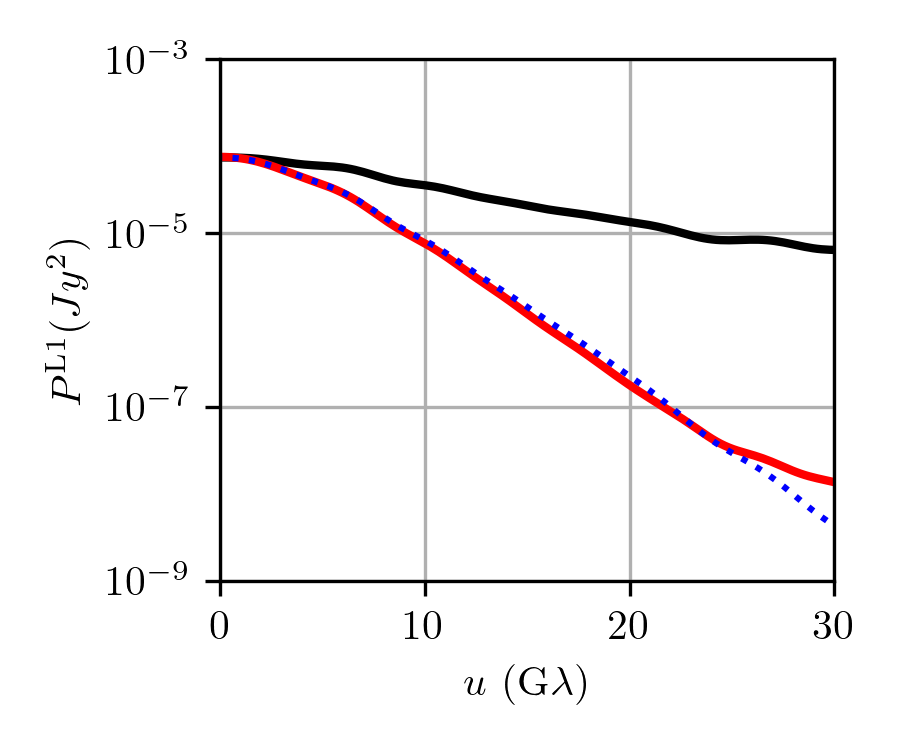}
\includegraphics[width=0.329\textwidth]{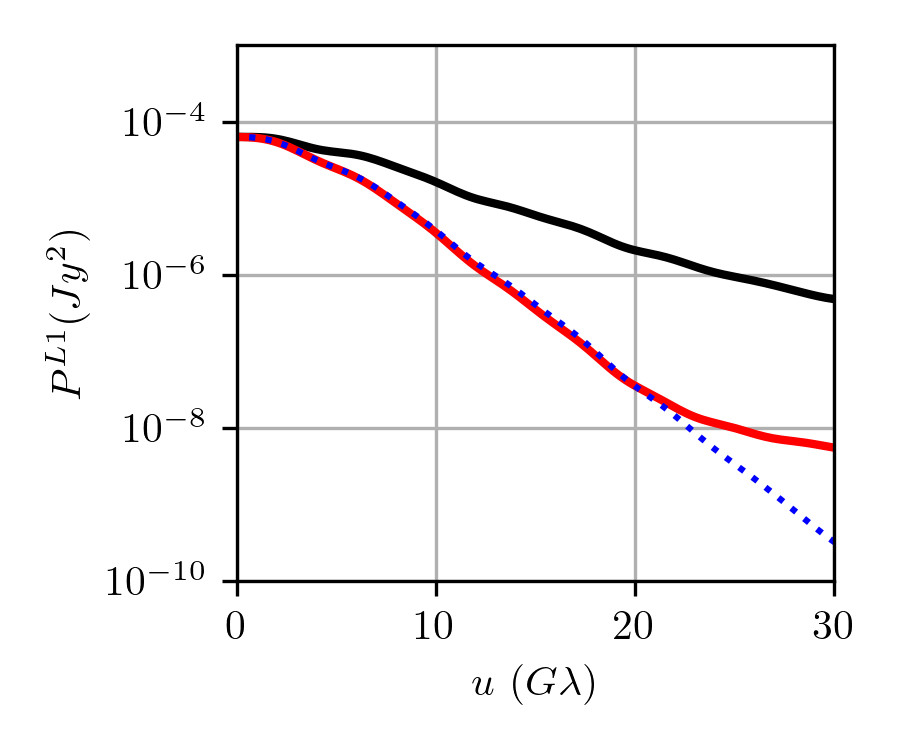}
\end{center}
\caption{Power spectra for the total intensity and two linear polarization modes. From left to right, the three panels are for total intensity, first linear polarization and second linear polarization. The black and red lines are the intrinsic and scattered power spectra, respectively. The blue dotted lines are the diffractively scattered power spectra. The green dashed line in the leftmost panel is the power spectrum of the Gaussian delta ring with radius of 25 $\muas$, which is the same envelope for the simple source model. 
}
\label{fig:PI,PL1,PL2} 
\end{figure*}

We first examine the impact of the scattering on the power spectra of the total intensity, which is shown in the left panel of \autoref{fig:PI,PL1,PL2}. The central peak is associated with the net source structure, i.e., a unit Jy Gaussian delta ring with radius of $25\,\muas$. The ringing is associated with the Fourier transform of the the delta ring as the envelope.
The extended plateau at baselines longer than $15\,\Gl$ is associated with two effects: first the small-scale variable structures that we seek to recover, and second the refractive scattering.

It is, in fact, the latter of these two that overwhelmingly dominates, as evidenced by the impact of diffractive scattering, shown by the dashed blue line, which strongly suppresses the contributions from the intrinsic source structure at $u\gtrsim10\,\Gl$. 


The power spectra associated with the polarization, shown in the center and right panels of \autoref{fig:PI,PL1,PL2}, do not have a prominent central peak because the total polarization flux vanishes on average (though while small, is non-zero for any given realization).  Thus, while again the diffractive scattering suppresses the power at long baselines due to intrinsic structure, there is much less contamination from refractive scattering, which dominates only for $u\gtrsim20\, \Gl$.

In all cases, the power spectra after scattering deviate substantially from those intrinsic to the source.  That is, as anticipated, due to both diffractive and refractive effects, the observed power spectra are themselves a poor proxy for the intrinsic power spectra, becoming worse as the spatial scale decreases.


\begin{figure*}
\begin{center}
\includegraphics[width=0.329\textwidth]{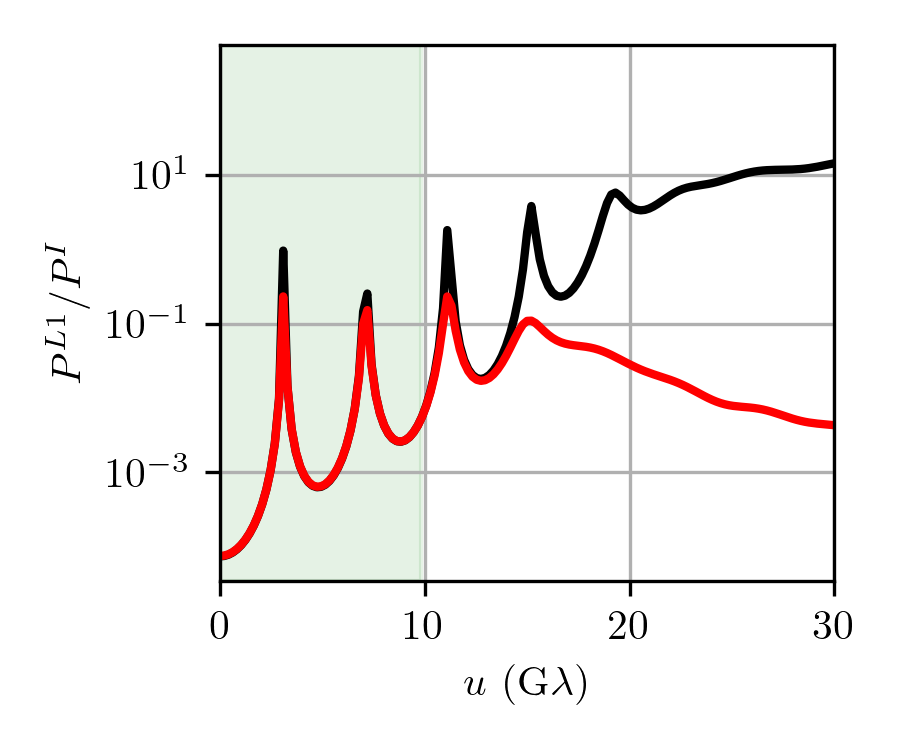}
\includegraphics[width=0.329\textwidth]{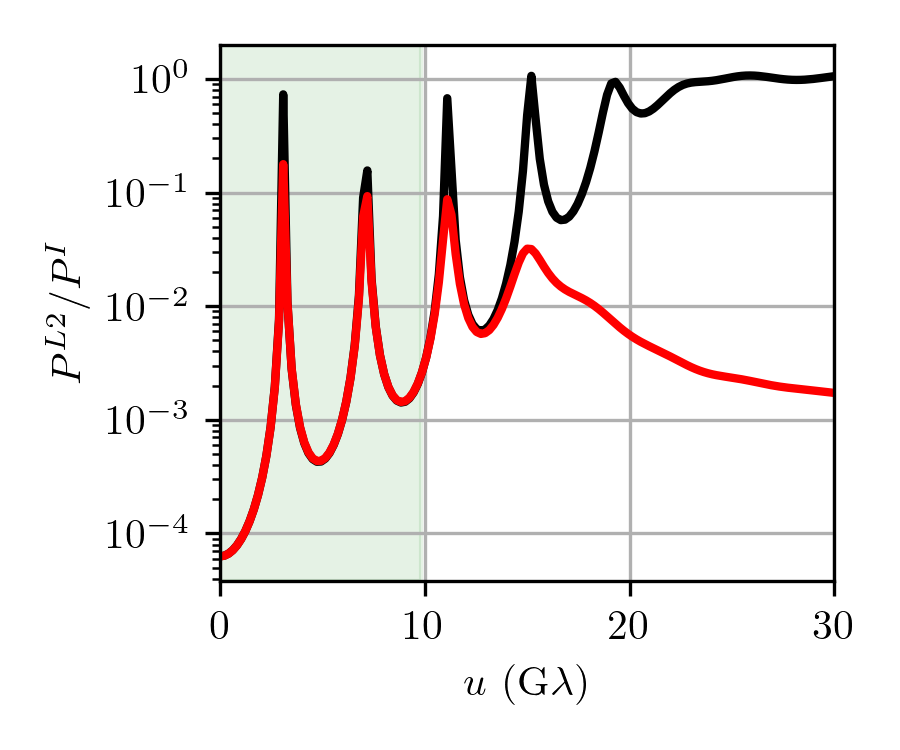}
\includegraphics[width=0.329\textwidth]{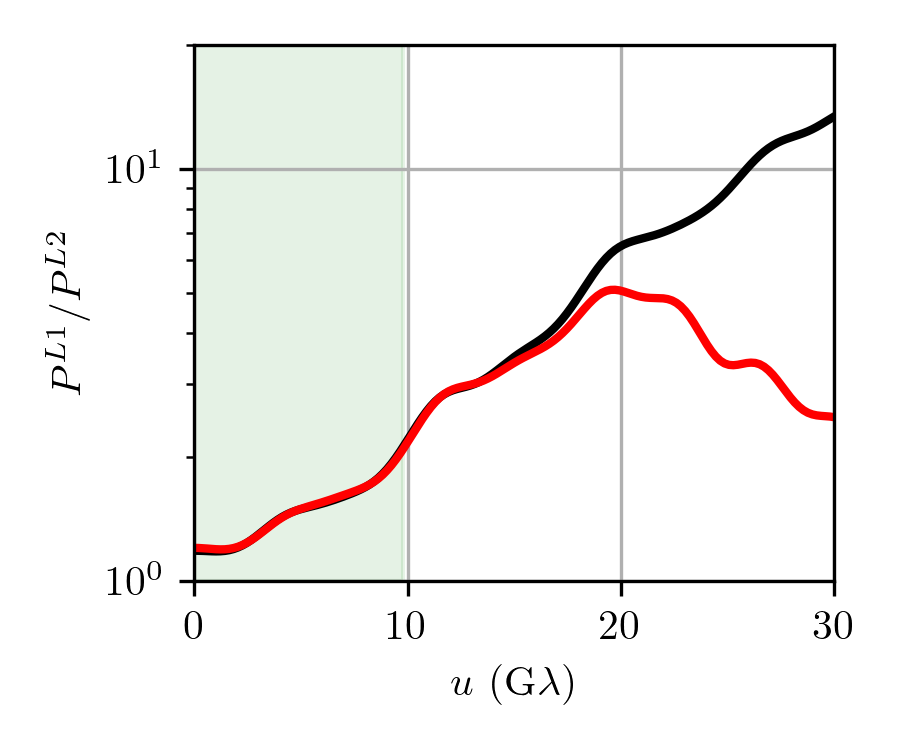}
\end{center}
\caption{Power spectra ratio between the total intensity and the first polarization mode (left), the total intensity and the second polarization mode (middle), and the two polarization modes (right) for the simple source structure defined in \autoref{sec:toy}. The black line shows the intrinsic power spectra ratio, while the red line is the power spectra ratio observed after scattering. The green band represents the size of Earth-bound baselines at 230 $\rm GHz$.}
\label{fig:L1I,L2I,L1L2} 
\end{figure*}

The ratios of observed power spectra, as defined in \autoref{eq:Pratio_approx} and shown in \autoref{fig:L1I,L2I,L1L2}, produce remarkable agreement with those from the intrinsic images (i.e., prior to scattering).  When the ratio is made with $P^I$, the intrinsic source structure and positive definite nature of the intensity responsible for the large bump in the left panel of \autoref{fig:PI,PL1,PL2} introduces a large depression, overwhelming any structure that may be attributed to the small-scale fluctuations. In addition, the dominance of refractive scattering at long baselines in $P^I$ results in a significant departure of the observed from the intrinsic power spectra ratios by $u=12\,\Gl$ as anticipated at the end of \autoref{sec:exp_Dphi}.

However, in stark contrast, the power spectra ratio of the two polarized modes, show in the right panel of \autoref{fig:L1I,L2I,L1L2}, match well out to $u\approx17\,\Gl$.  While beyond $u\approx20\,\Gl$, refractive scattering drives large deviations from intrinsic power spectra ratio, this is well beyond the baselines accessible to EHT and ngEHT at 1.3\,mm.  

Similar experiments were performed for a variety of choices of the $\alpha$, $\alpha_{L1}$, and $\alpha_{L2}$, with similar results. 

While the above is schematic, involving only a very simple source structure, nevertheless a number of immediate conclusions can be drawn that we will see reflected in the more physically applicable demonstrations that follow:
\begin{itemize}
    \item Observed power spectra are poor estimators in the presence of interstellar scattering for the intrinsic variable structures.
    \item For all polarization modes, diffractive scattering suppresses long-baseline observed power spectra.
    \item For polarization modes with large net flux (e.g., Stokes $I$), refractive scattering substantially contaminates the power spectra.
    \item Ratios of power spectra generally produced better estimates of the corresponding intrinsic quantities.
    \item Ratios of power spectra associated with polarization modes with zero net flux are substantially more accurate estimates, extending well beyond the spatial frequency range accessible from the ground at 1.3\,mm.
\end{itemize}
Based on the above, we conclude that when the polarization modes are well chosen, the approximation in \autoref{eq:Pratio_approx} is well motivated.

\section{validation with GRMHD simulations} \label{sec:GRMHD}

In contrast to the simple, phenomenological models discussed in the previous section, GRMHD simulations provide a natural astrophysically-motivated set of complex source stuctures.
While GRMHD simulations do not afford the freedom to arbitrarily modify the input fluctuation spectra, they do incorporate credible realizations of the anticipated  turbulence  and  magnitude  of  the  polarized flux.  As a result, it is possible to reasonably assess the practical limitations imposed by thermal noise noise and limited number of observations to be averaged. Here we repeat the kinds of tests performed in Section \ref{sec:toy} for one such GRMHD simulation, taken from the set presented in \citet{PaperV}.

\subsection{GRMHD Simulated Intrinsic Image and Simulated Observation} \label{GRMHDIntrinsic}

We employ a SANE, $a=0$, $i=30^\circ$, $R_{\rm high}=40$ simulation from the set presented in \citet{PaperV}, to which we direct the reader for information about the simulation particulars.  For our purposes, it is important only that the simulation presents a physically self-consistent realization of the kind of turbulence, degree of net polarization ($\sim$3\%), and typical polarization fractions ($\sim$20\%) appropriate for Sgr A*. 

\begin{figure}
\begin{center}
\includegraphics[width=\columnwidth]{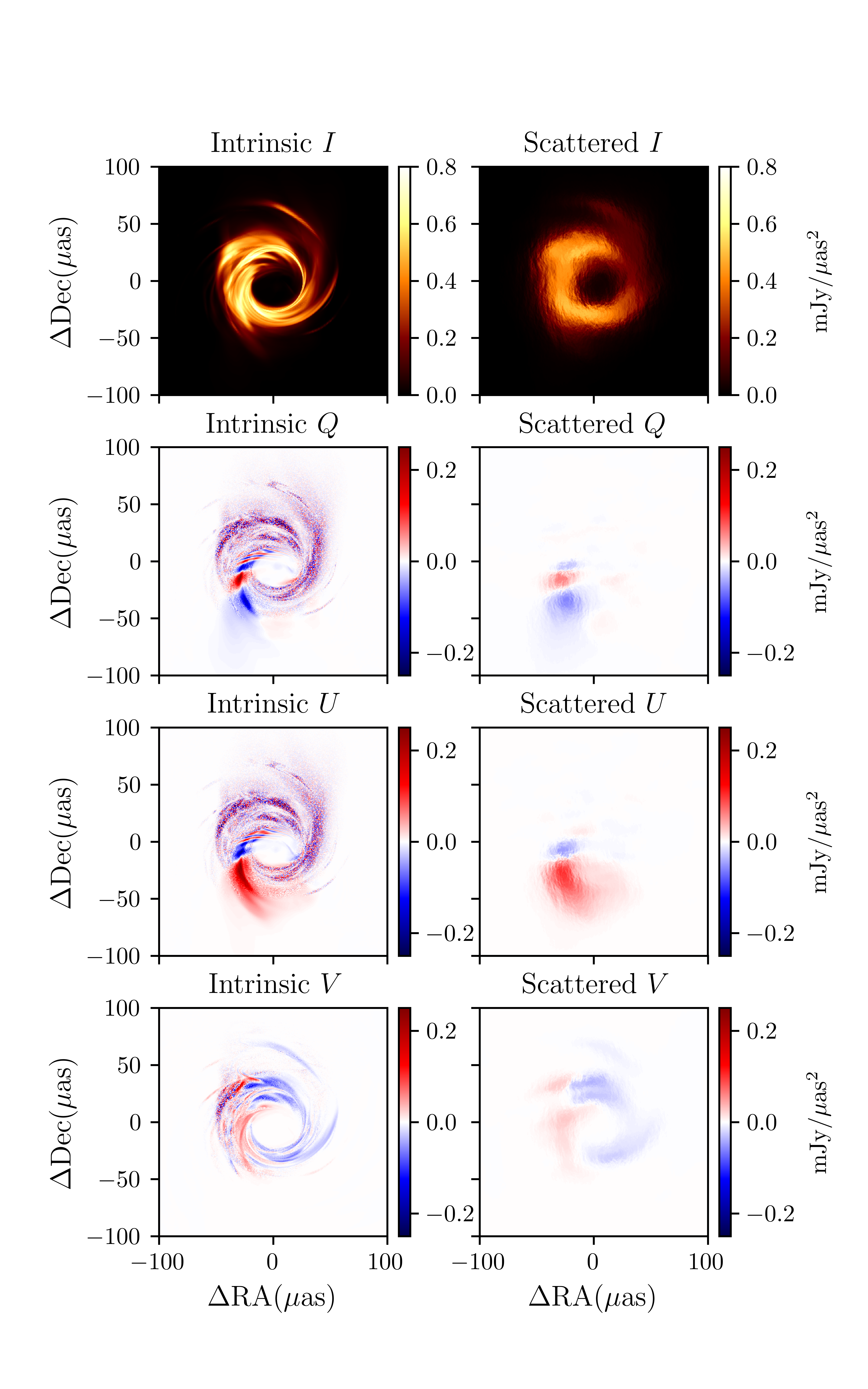}
\end{center}
\caption{Arbitrary snapshot drawn from the GRMHD simulation used. The figure includes the Stokes I, Q, U and V components (left) and their corresponding scattered versions (right). Bottom row: the two constructed polarization modes, $S1$ and $S2$, are shown, both intrinsic and scattered.
}
\label{fig:GRMHDsnapshots} 
\end{figure}

\begin{figure}
\begin{center}
\includegraphics[width=\columnwidth]{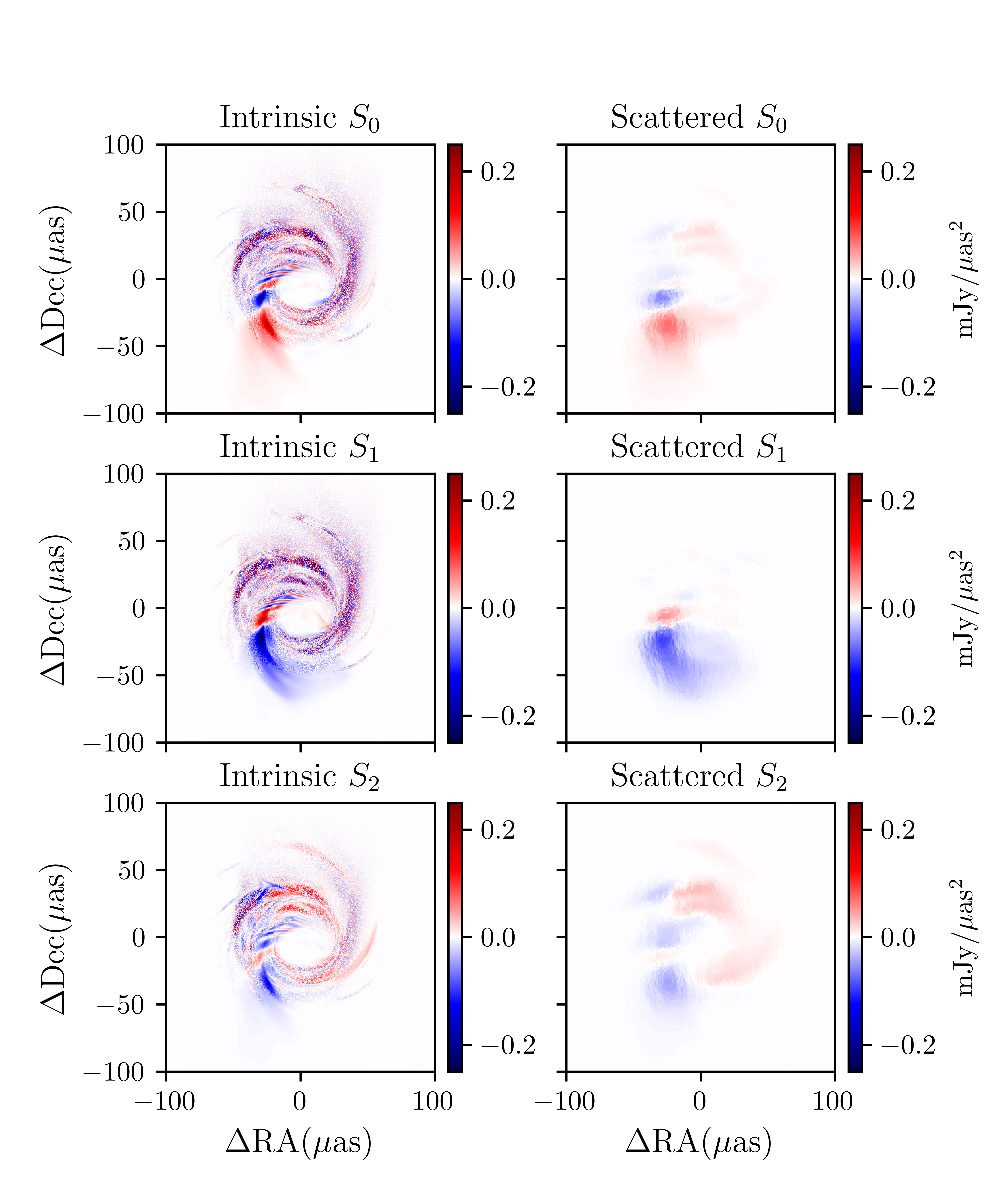}
\end{center}
\caption{The same arbitrary snapshot drawn from the GRMHD simulation used in \autoref{fig:GRMHDsnapshots}. The figure includes the two constructed polarization modes, $S1$ and $S2$, both intrinsic and scattered.
}
\label{fig:GRMHDsnapshots_S1S2} 
\end{figure}

An arbitrary snapshot drawn from the simulation is shown in \autoref{fig:GRMHDsnapshots} with its four Stokes components, and in \autoref{fig:GRMHDsnapshots_S1S2} with the constructed polarization modes $\textbf{S}_0$, $\textbf{S}_1$ and $\textbf{S}_2$, as stated in \autoref{subsubsec:ratio}.

Note that because our goal here is not to predict the statistics of GRMHD simulation images, but rather to demonstrate the ability to faithfully retrieve statistical elements of the underlying intrinsic images in the presence of scattering, this single GRMHD simulation is sufficient for our purposes.

The simulated emission is assumed to arise from synchrotron emission due to a population of hot electrons. The simulation data contains the total intensity and polarization maps at 1.3~mm for accretion flow parameters relevant for Sgr A*, i.e., the four Stokes parameters, $I$, $Q$, $U$ and $V$, for 3000 individual snapshots \citep[][and references therein]{PaperV}.  
From these intrinsic images, scattered images are produced using the Stochastic Optics package within \ehtim in manner identical to that used in \autoref{sec:toy}.



We follow a nearly identical procedure to generate simulated observations from the GRMHD simulations as that described in \autoref{subsec:ToyModelObs}.  This procedure differs in that the polarization maps are now identified with the three polarized Stokes maps ($Q$, $U$, $V$).  Where these are scattered, we produce a new scattering screen realization for each image. 
For each frame, we construct the single-snapshot estimate of the spatial power spectra associated with $I$, $Q$, $U$, and $V$ are constructed, i.e., $\abs{V^{I}}^2$, $\abs{V^{Q}}^2$, $\abs{V^{U}}^2$, and $\abs{V^{V}}^2$, respectively.  This procedure is repeated for every image in the GRMHD simulation to generate estimates for the ensemble- and turbulence-averaged estimates of the spatial power spectra.

\subsection{Spatial Power Uncertainty Estimate} \label{subsec:ErrorEstimate}

To assess if the observed and intrinsic spatial power spectra are distinguishable, we require an estimate of the anticipated uncertainty on the spatial power spectra.  This arises from multiple potential origins.  However, there are two irreducible contributors: thermal noise associated with the individual stations within the EHT and ngEHT, and the sampling uncertainty due to an insufficiently complete ensemble.  Here we describe how we estimate each of these.  Note that we make aggressive assumptions to {\em reduce} both sources of uncertainty, thereby enforcing a {\em stricter} limit on the required fidelity of the spatial power spectra ratios.

\begin{figure*}
\begin{center}
\includegraphics[width=0.329\textwidth]{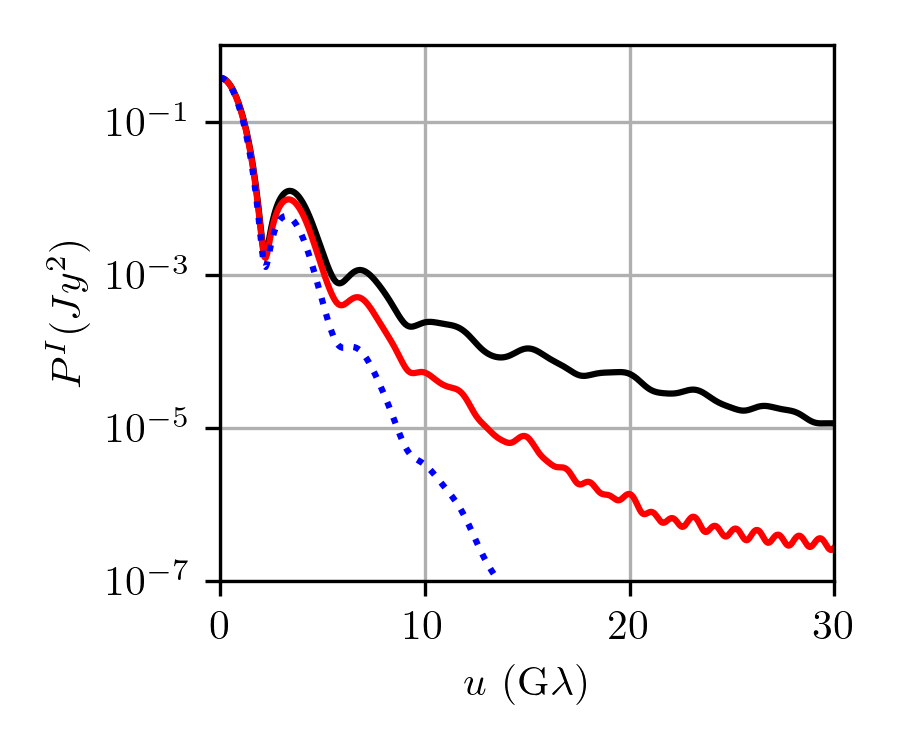}
\includegraphics[width=0.329\textwidth]{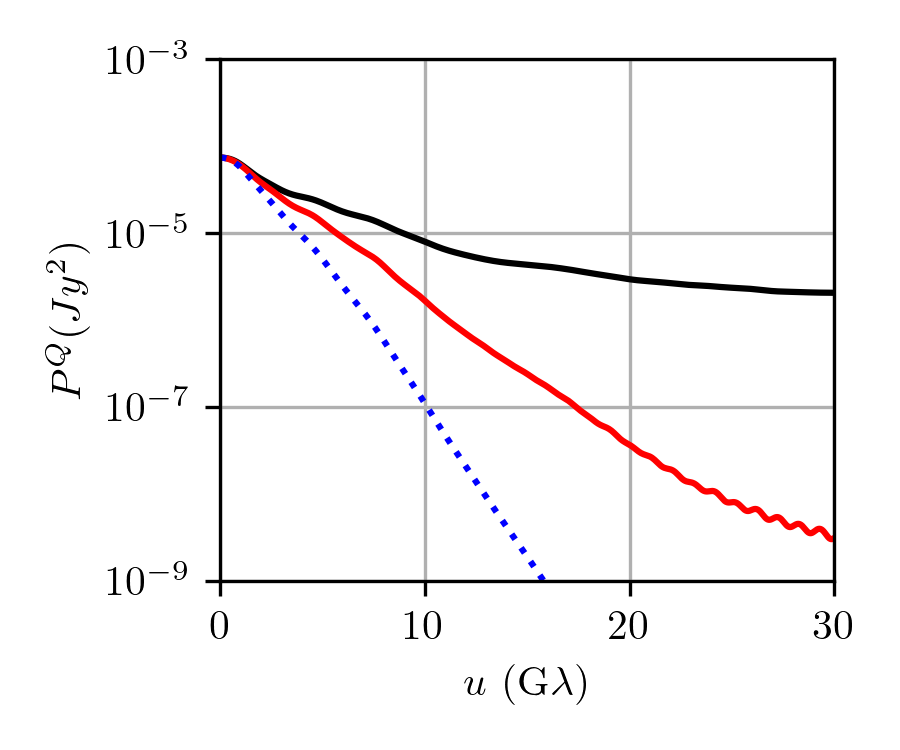}
\includegraphics[width=0.329\textwidth]{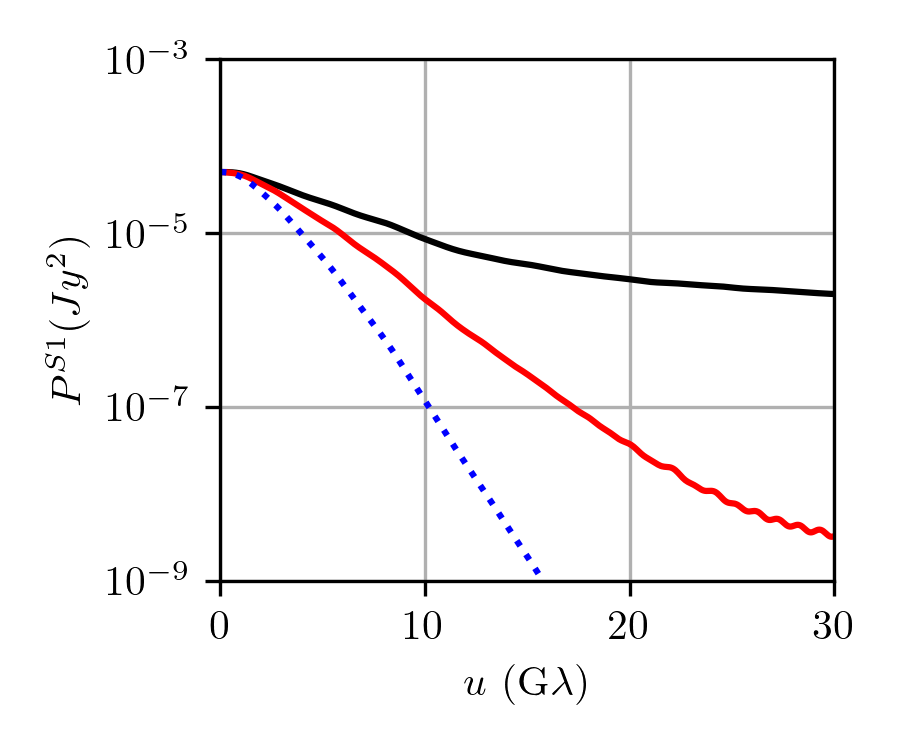}
\caption{Power spectra for Stokes I (left), Stokes Q (middle), and the optimal $S1$ (right) before (black) and after (red) application of the scattering screen. The dotted blue line represents the diffractively scattered power spectrum}.
\label{fig:GRMHD,PI,PQ,PS1}
\end{center}
\end{figure*}

\subsubsection{Thermal Noise Estimates} \label{thermal}

The thermal error at EHT stations arises from a number of potential sources, including the atmosphere, side lobes picking up the local environment, and the electronics within receiver.  The combination is typically characterized by a system equivalent flux density, SEFD.  In terms of these, the uncertainty on a visibility measured by stations $A$ and $B$ with a bandwidth $\Delta B$ and coherently averaged over a time $\tau$,
\begin{equation}
\sigma_{AB} = \sqrt{\frac{{\rm SEFD}_A {\rm SEFD}_B}{2 \tau \Delta B}},
\end{equation}

SEFDs for EHT stations are listed in Table 2 of \citep{2019ApJ...875L...3E}.  These range from 74~Jy for ALMA to 19300~Jy for SPT.  We adopt, for illustration, PV and APEX, for which the SEFDs are 1900~Jy and 4700~Jy, respectively, the intermediate SEFDs in the EHT.  Were we to adopt the two stations with the highest SEFDs in the EHT, SMT and SPT, the estimated thermal noise could be a factor of 30 higher. The median thermal noise across the EHT is roughly an order of magnitude larger than the minimal value.  Finally, we adopt a bandwidth of $4~{\rm GHz}$, corresponding to the combination of high- and low-band data from the EHT, and an integration time of $10~{\rm min}$, corresponding to a typical scan time, yielding $\sigma_{\rm th}\approx0.4~{\rm mJy}$.  Upon averaging $N$ independent observations, the effect thermal noise is reduced by a further factor of $N^{-1/2}$.  Note that this is independent of the particular polarization mode under consideration.

The thermal noise on the spatial power spectra after averaging $N$ independent observations is obtained via standard error propagation,
\begin{equation}
\sigma_{\langle V_A^2\rangle}
=
\frac{2\sigma_{\rm th}}{N^{1/2}}
\left<\abs{V_A}^2\right>^{1/2}.
\end{equation}
The thermal noise on the ratio of spatial power spectra after averaging $N$ observations is, 
\begin{multline}
\sigma_{\langle V_A^2\rangle/\langle V_B^2\rangle}
=\\
\frac{2\sigma_{\rm th}}{N^{1/2}}
\frac{\left<\abs{V_A}^2\right>}{\left<\abs{V_B}^2\right>}
\left(
\frac{1}{\left<\abs{V_A}^2\right>}
+
\frac{1}{\left<\abs{V_B}^2\right>}
\right)^{1/2}.
\end{multline}

\subsubsection{Sampling Noise Estimates} \label{ensemble}

The sampling noise describes the uncertainty associated with having a finite number of samples in the estimate of the ensemble and turbulence averages.  When the number of samples, $N$, is large, the central limit theorem implies that this is related to the variance of the visibility amplitude, i.e.,
\begin{equation}
\Sigma_{V_A} = \left[\left<\abs{V_A}^2\right> - \left<\abs{V_A}\right>^2 \right]^{1/2}.
\end{equation}
Note that unlike the thermal noise, this differs between the various Stokes parameters, which may exhibit different degrees of variability.
From $\Sigma_{\langle V_A\rangle}$, the uncertainties on the spatial power spectrum and spatial power spectra ratios can be immediately constructed via the standard error propagation,
\begin{equation}
\Sigma_{\langle V_A^2 \rangle}
=
\frac{2\Sigma_{\langle V_A\rangle}}{N^{1/2}}
\left<\abs{V_A}^2\right>^{1/2},
\end{equation}
and
\begin{multline}
\Sigma_{\langle V_A^2\rangle/\langle V_B^2\rangle}
=\\
\frac{2}{N^{1/2}}
\frac{\left<\abs{V_A}^2\right>}{\left<\abs{V_B}^2\right>}
\left(
\frac{\Sigma_{\langle V_A\rangle}^2}{\left<\abs{V_A}^2\right>}
+
\frac{\Sigma_{\langle V_B\rangle}^2}{\left<\abs{V_B}^2\right>}
\right)^{1/2}.
\end{multline}

\subsection{Power Spectra Estimation Results}

\begin{figure*}
\begin{center}
\includegraphics[width=0.329\textwidth]{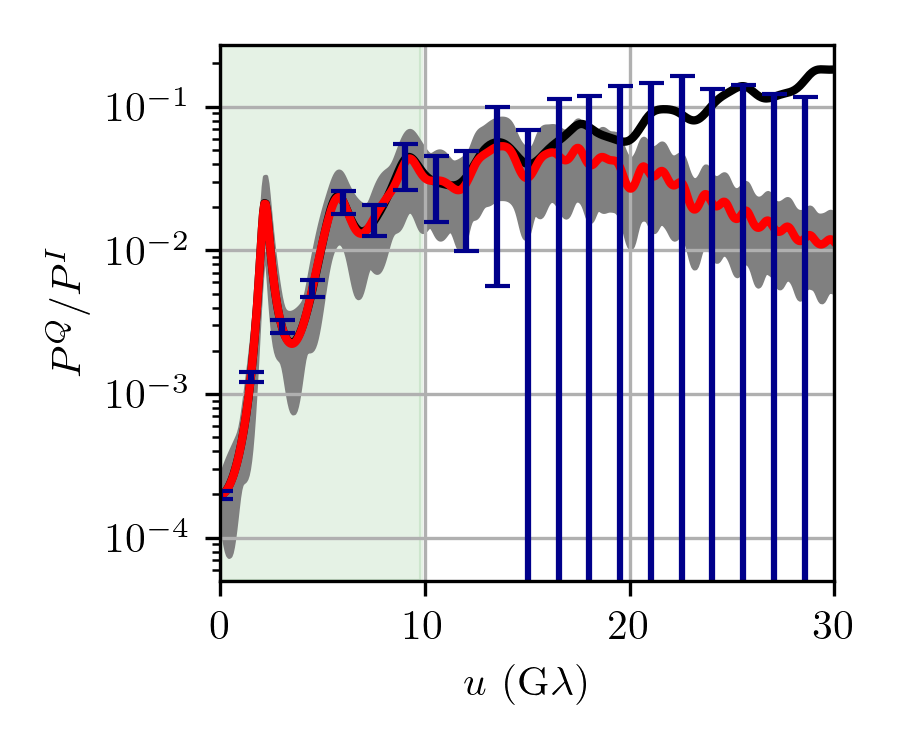}
\includegraphics[width=0.329\textwidth]{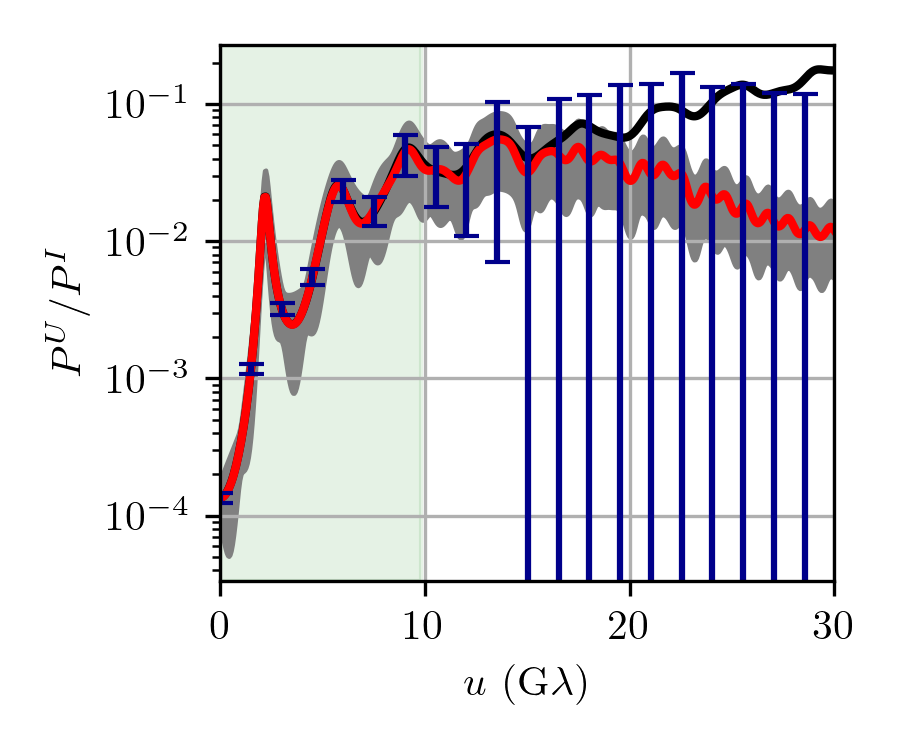}
\includegraphics[width=0.329\textwidth]{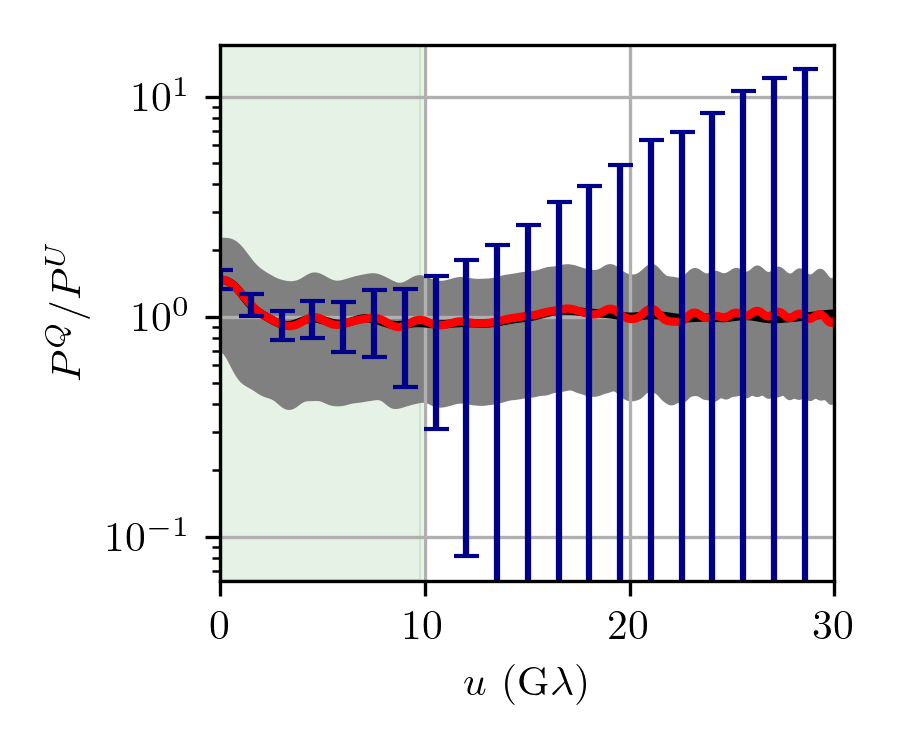}
\caption{Ratios of the power spectra of different polarization modes. From left to right, the three panels show the power spectra ratio for $P^Q/P^I$, $P^U/P^I$ and $P^Q/P^U$. The black and red line represents before and after the imposition of scattering. The blue error bars are the thermal noise associated with telescopes. We used PV and APEX, which have the intermediate sensitivities among all EHT telescopes, to generate the error bars. The grey error bands are the sampling noise associated with the intrinsic spatial variability of the intrinsic source, which we averaged down assuming 25 independent observations. The green band represents the size of Earth-bound baselines at 230 $\rm GHz$.}
\label{fig:GRMHDratio}
\end{center}
\end{figure*}

\begin{figure*}
\includegraphics[width=0.33\textwidth]{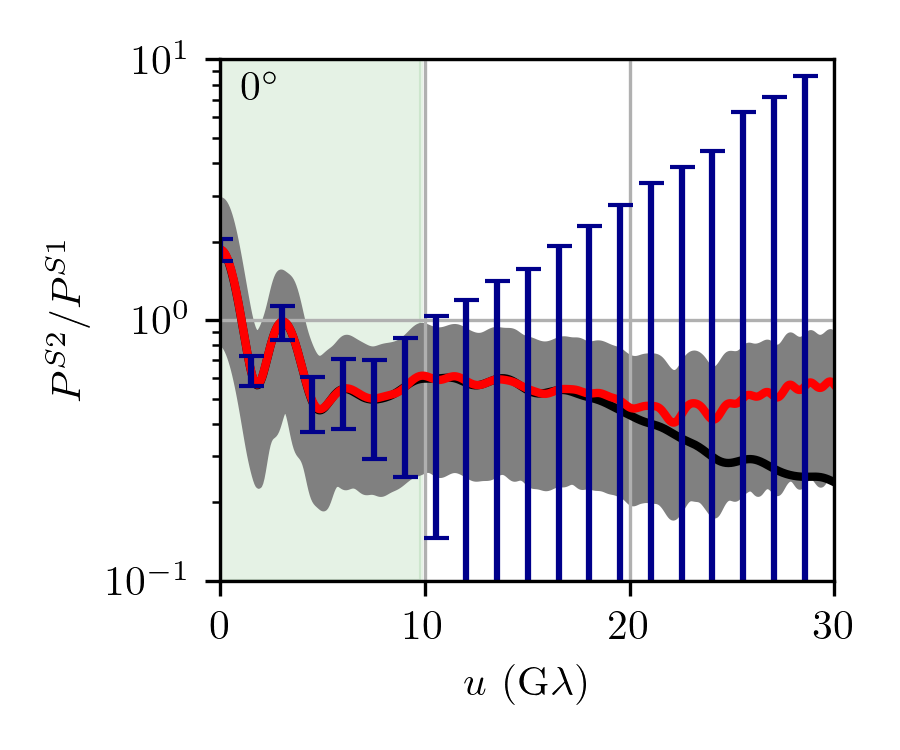}\hfill
\includegraphics[width=0.33\textwidth]{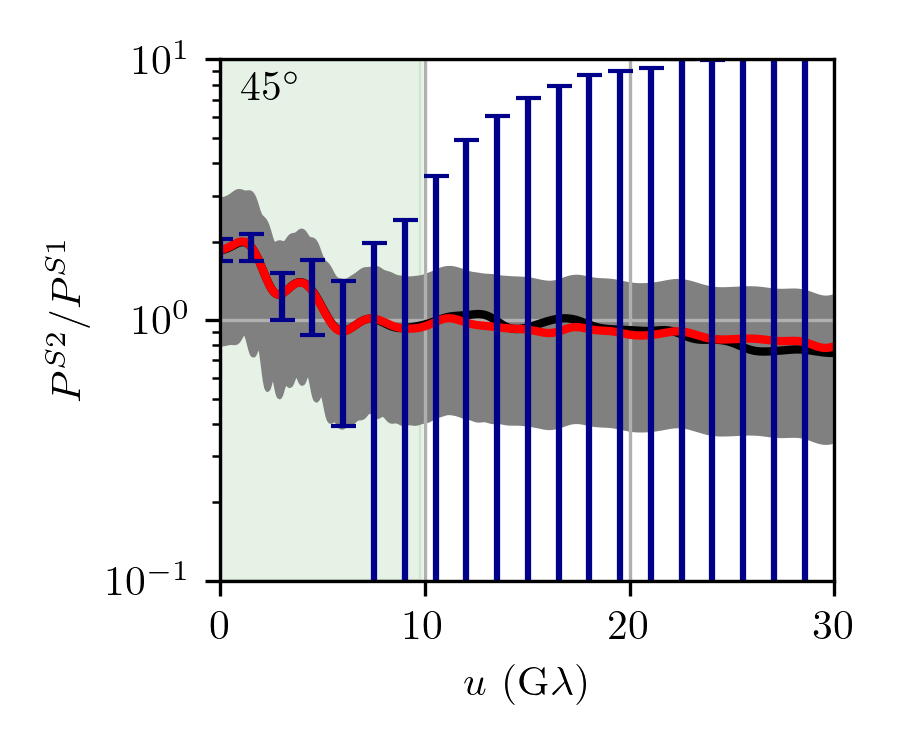}\hfill
\includegraphics[width=0.33\textwidth]{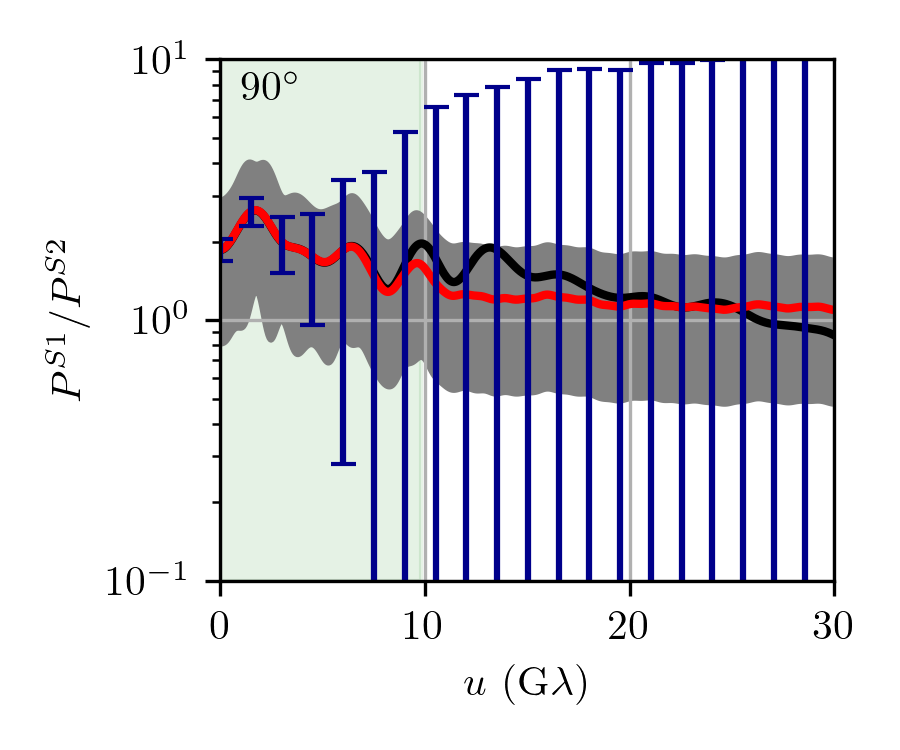}
\caption{Intrinsic (black) and observed (red) $P^{S2}/p^{S1}$ along rays at different orientations relative to the $u$ axis.  From left to right: $0^\circ$ ($u$ axis), $45^\circ$, and $90^\circ$ ($v$ axis). The sampling noise and thermal errors associated with $N=25$ observations are indicated by the grey band and blue error bars, respectively.}
\label{fig:GRMHDratioDifferentAngles}
\end{figure*}

The mean power spectra associated with Stokes $I$, $Q$, and $S1$ are shown in \autoref{fig:GRMHD,PI,PQ,PS1}.  As with the toy model presented in \autoref{sec:toy}, the non-zero total flux results in a peak at $u=0\,\Gl$, and a deficit associated with the diffractive component of the scattering at long baselines.  As with \autoref{fig:PI,PL1,PL2}, refractive scattering lessens the reduction, seen most prominently in $P^I$ because of the comparatively large net value. In all cases, scattering significantly suppresses the power spectra relative to their intrinsic values, as anticipated.

The mean $P^Q/P^I$ and $P^{S2}/P^{S1}$ power spectra ratios are shown in \autoref{fig:GRMHDratio}, and are directly comparable to those in \autoref{fig:L1I,L2I,L1L2}.  In addition, the sampling and thermal error scales  for $N=25$ independent observations of the source and scattering screen are indicated, providing a natural assessment of the accuracy of the approximation in \autoref{eq:Pratio_approx}.  Apart from differences in the underlying source structure, e.g., the clearly evident oscillations associated with the lensed emission ring, the gross properties noted in \autoref{sec:toy} remain: the suppression at short baselines by the non-zero mean total flux in the $P^Q/P^I$ ratio (similar to $P^U/P^I$), and lack of such a suppression in the Stokes basis defined by $\textbf{S}_1$ and $\textbf{S}_2$.  Even at the longest ground-based baseline lengths -- $10\,\Gl$ at 230\,GHz -- the impact of the scattering screen is effectively mitigated.

\begin{figure*}
\includegraphics[width=0.329\textwidth]{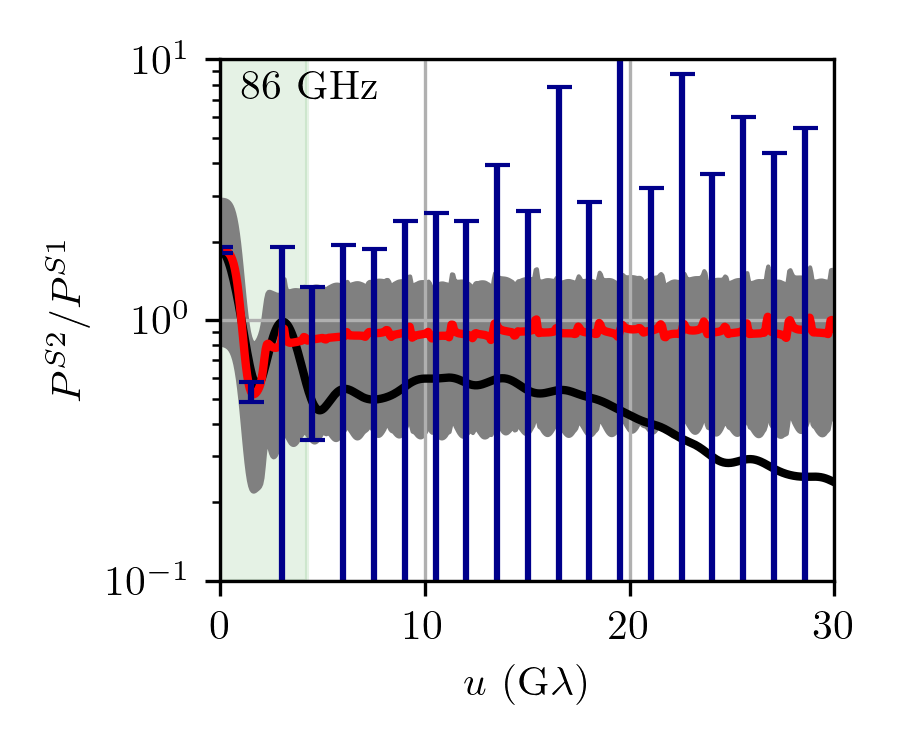}
\includegraphics[width=0.329\textwidth]{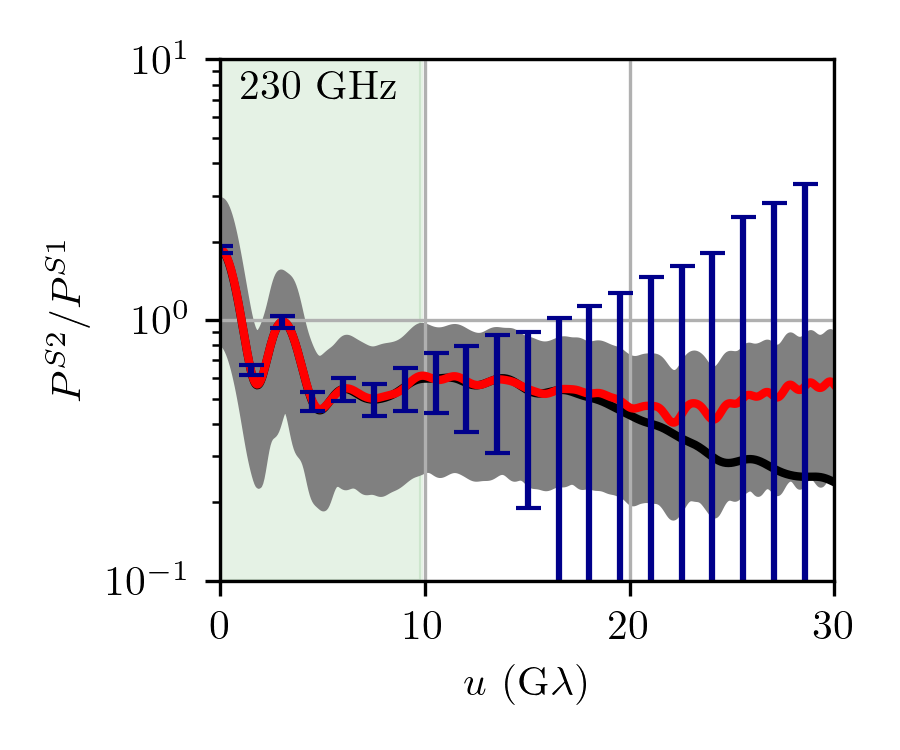}
\includegraphics[width=0.329\textwidth]{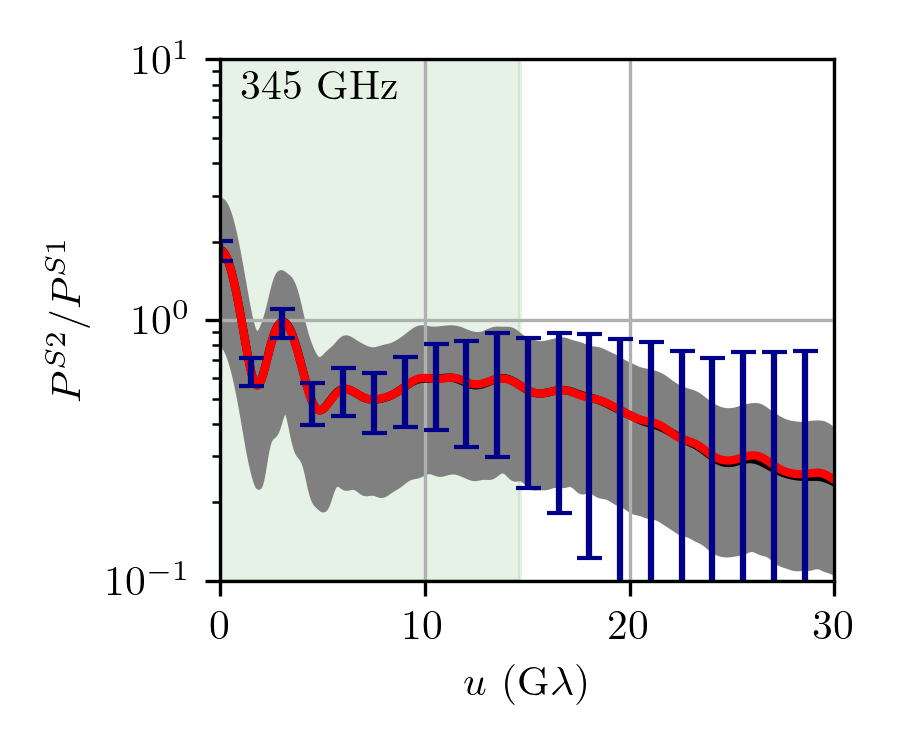}
\caption{Mean intrinsic (black) and observed (red) ratio of the power spectra associated with polarization modes defined by the Stokes vectors $\textbf{S}_1$ and $\textbf{S}_2$ at 86\,GHz (left), 230\,GHz (middel) and 345\,GHz (right).  The sampling noise and thermal errors associated with $N=25$ observations and based on ngEHT telescope sensitivities are indicated by the grey band and blue error bars, respectively. The green band represents the size of Earth-bound baselines at 86 $\rm GHz$, 230 $\rm GHz$ and 345 $\rm GHz$, from left to right.}
\label{fig:GRMHDratio_345}
\end{figure*}

While \autoref{fig:GRMHDratio} shows only the ray through the uv-plane along the $u$-axis, this improvement is generic.  \autoref{fig:GRMHDratioDifferentAngles} presents $P^{S2}/P^{S1}$ for radial rays at different orientations.  While the magnitude of the discrepancies between the observed and intrinsic power spectra ratios and the location where they begin to differ varies, in all cases, at all baselines relevant for EHT and ngEHT, these discrepancies are small in comparison to relevant uncertainties.

In addition to vastly increasing the number and density of baselines available, the ngEHT envisions receivers with increased bandwidth and improved detector efficiency that will improve sensitivity across the array, reducing the thermal noise contributions to the visibility uncertainties \citep{ngEHTsites}.  For nominal SEFDs associated with the reference ngEHT array in \citet{ngEHTsites} and a bandwidth of 16\,GHz, we show $P^{S2}/P^{S1}$ at the observing frequencies under discussion for the ngEHT in \autoref{fig:GRMHDratio_345}.
As shown in the middle panel of \autoref{fig:GRMHDratio_345}, for the ngEHT, thermal noise ceases to be the chief impediment to measuring the intrinsic power spectra ratios below $\sim15\,\Gl$, covering all Earth-sized baselines.
The right panel of \autoref{fig:GRMHDratio_345} shows $P^{S2}/P^{S1}$ at 345\,GHz with ngEHT telescope sensitivities and bandwidths; again all ground-base baselines are dominated by the sampling uncertainty associated with source variability.  Out to $\sim50\,\Gl$, nearly twice the $30\,\Gl$ region shown in \autoref{fig:GRMHDratio_345}, the observed mean power spectra ratio provides an excellent estimate of the intrinsic ratio.  Within $\sim100\,\Gl$ observed and intrinsic power spectra ratio differ by less than the sampling uncertainty.  These imply that future high-frequency polarimetric observations by EHT, ngEHT, and space-based mm-VLBI experiments will all be able to accurately statistically probe the turbulent structures in horizon-scale targets on scales from $7\,\muas$ (ground) to $2\,\muas$ (space). 
Whereas at lower observation frequencies, the impact of scattering is enhanced. In the left panel of \autoref{fig:GRMHDratio_345}, which is at 86\,GHz, although scattering deviates the power spectrum ratio from the intrinsic at as early as $3\,\Gl$, which is the Earth-based VLBI baseline limit at this frequency, they are within the allowed sampling uncertainty until $20\,\Gl$. This implies the possibility to apply this scattering mitigation scheme to data from GMVA and other telescopes with lower observation frequencies.


\begin{figure}
\includegraphics[width=\columnwidth]{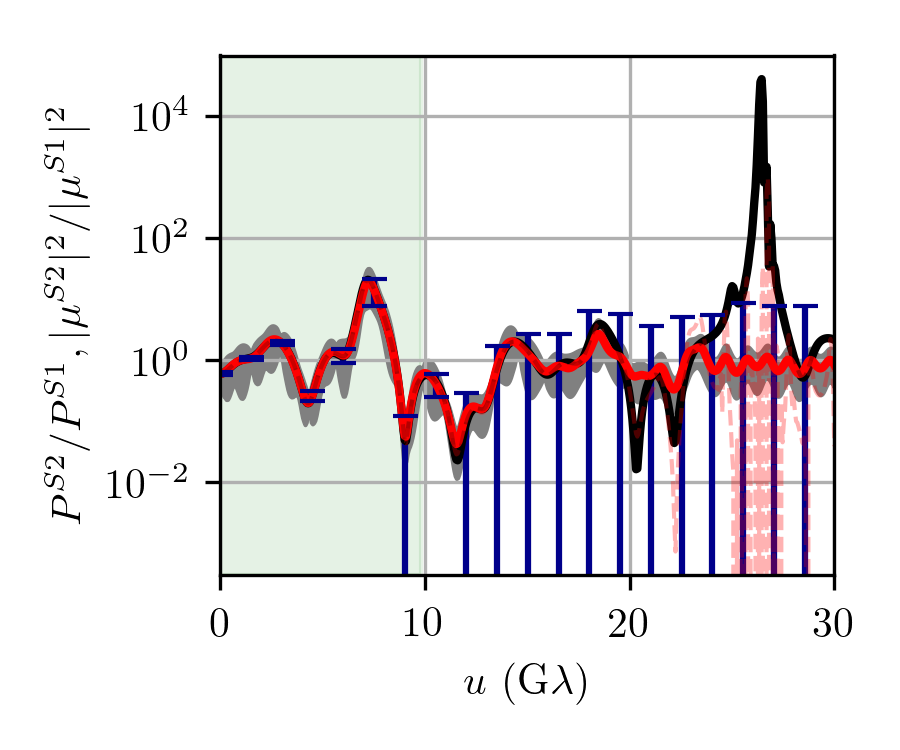}
\includegraphics[width=\columnwidth]{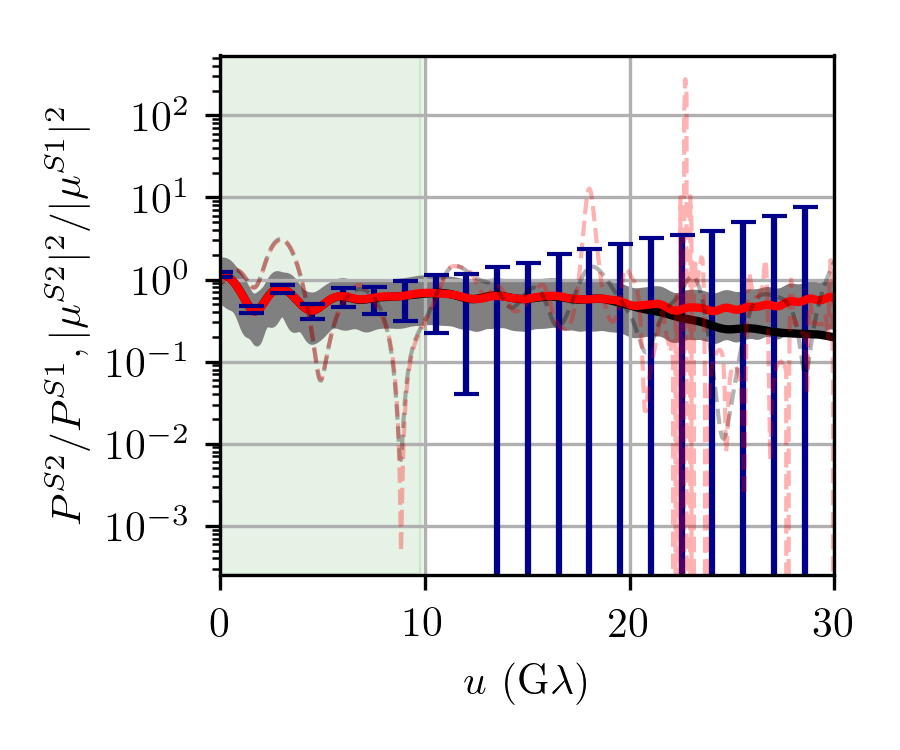}
\caption{Ratios of the power spectra (solid lines) and means (dashed lines) for the two constructed polarization modes, $\textbf{S}_1$ and $\textbf{S}_2$. Top: a single, static intrinsic image, represented by a single GRMHD snapshot, viewed through an evolving scattering screen.  Bottom: an evolving intrinsic image and scattering screen.  In both panels, the sampling noise and thermal errors associated with N = 25 observations and based on ngEHT telescope sensitivities are indicated by the grey band and blue error bars, respectively. The green band represents the size of Earth-bound baselines at 230 GHz.}
\label{fig:static_dyanmic_ratio}
\end{figure}

The ability explicitly distinguish between intrinsic source variability and extrinsic evolution in the scattering screen using power spectra and mean ratios is demonstrated in \autoref{fig:static_dyanmic_ratio}.  Like $P^{S2}/P^{S1}$, the ratio of the means, $|\mu^{S2}|^2/|\mu^{S1}|^2$, are insensitive to the scattering for baselines relevant for current and future Earth-bound VLBI arrays.  When the source is static, here taken to be a single GRMHD snapshot, the two sets of ratios are identical out to baselines well in excess to those accessible on the ground, ultimately limited by diffractive scattering. In contrast, for variable sources, the two ratios differ at high significance.  Therefore, the comparison of these two ratios, i.e., $P^{S2}/P^{S1}$ and $|\mu^{S2}|^2/|\mu^{S1}|^2$, presents a way in which to find direct evidence for intrinsic source evolution.



Which component of the uncertainty dominates depends on baseline length.  At sufficiently large $u$ the strong suppression due to diffractive scattering reduces the signal precipitously. The lower intrinsic $S/N$ of the polarized data result in characteristically larger thermal uncertainties on the power spectra ratio.  At short baselines, the sampling error dominates.  These two regimes conceptually differ in the manner that measurements can be improved.  The thermal noise can be reduced by improvements to station sensitivity (e.g., increased bandwidth, dish size, phase referencing, etc.).  In contrast, the sampling noise is a consequence of the intrinsic variability alone, and can only be improved by repeated observation.  Because additional observation epochs also reduce the thermal noise at the same rate, where the sampling noise dominates, it will do so regardless of the number of observations.

\begin{figure}
\includegraphics[width=\columnwidth]{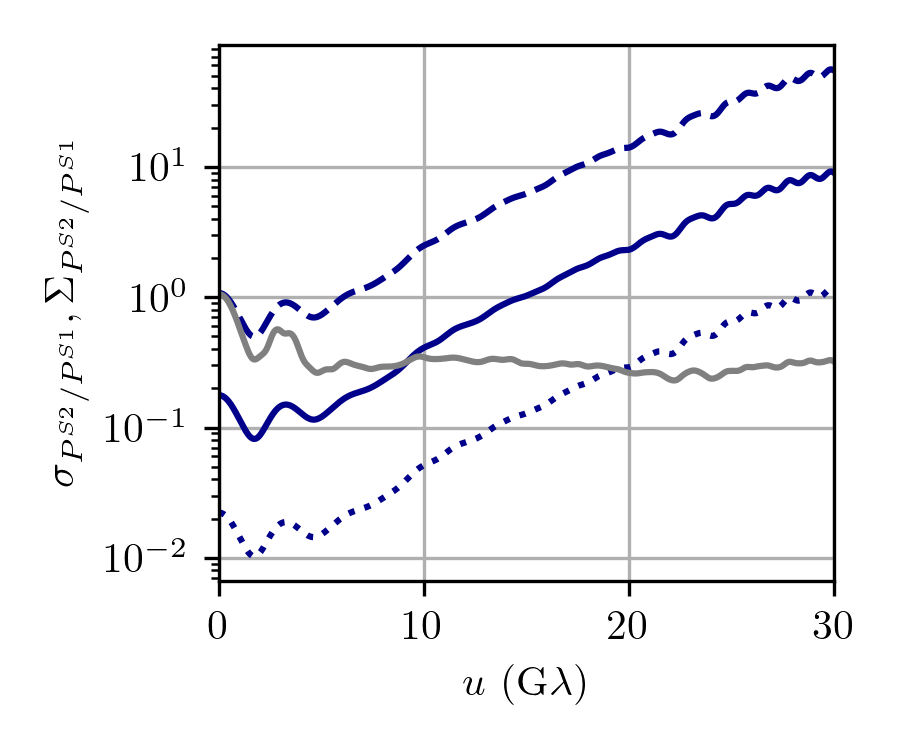}
\caption{Comparison of the sampling and thermal noise estimates for $P^{S2}/P^{S1}$ at 230\,GHz.  The thermal noise for the minimum, median, and maximum noise estimates are shown by the dotted, solid, and dashed blue lines.  The sampling noise is indicated by the grey line.}
\label{fig:GRMHDratio_noise}
\end{figure}

The location of the transition from sampling dominated to thermally dominated noise depends on the nature of the variability and the sensitivity of the individual stations (see \autoref{fig:GRMHDratio_noise}).  For the simulation considered here and the median thermal noise, this transition occurs just beyond $10\,\Gl$.  However, for the baseline with the maximum thermal noise, the uncertainty at all baselines is dominated by the thermal component, implying that sensitivity of existing EHT stations is likely to be the limiting factor in the accuracy with which the power spectra ratios can be measured.

\section{Conclusions} \label{sec:C}

Interstellar scattering associated with turbulence in the ISM is nonbirefringent for physically reasonable magnetic field strengths.  As a consequence, the effect of scattering on horizon-resolving polarization maps of Sgr A* is expected to be independent of the polarization mode being observed.  This presents an opportunity to statistically separate the small-scale structures induced by refractive scattering and those intrinsic to the source, presumably due to turbulence within the near-horizon emission region.

We characterize the statistical properties of the structural variability by their power spectra, $P^S$, defined to be the mean squared visibility associated with Stokes parameter $S$.  This definition is convenient because the impact of scattering is a linear operator that may be expressed as a tensor convolution acting on the $P^S$.  As far as $S/N$ permits, this convolution may be inverted via a perturbative expansion; for existing and proposed Earth-sized mm-VLBI arrays, like the EHT and ngEHT, only the first term in the expansion is required.  This effectively reduces to applying constraints to the ratio of $P^S$, and is otherwise insensitive to the details of the scattering screen.

It is possible to select the polarization modes to minimize the impact of refractive scattering and simplify the interpretation of the $P^S$.  We do this by constructing a particular basis of Stokes vectors, $\textbf{S}_1$ and $\textbf{S}_2$, that are orthogonal to the source-integrated mean Stokes vector.  These correspond to the fluctuations in the projected orientation of the magnetic fields on the sky and along the line of sight, respectively.  Thus, the ratio $P^{S2}/P^{S1}$ is a direct measure of the degree of isotropy in the MHD turbulence across spatial scale.

Using both a toy model, in which there is substantial control over the properties of the turbulent structures, and a GRMHD simulation, which contains a realistic representation of MHD turbulence, we have demonstrated the ability to reconstruct various intrinsic power spectra ratios, including $P^{S2}/P^{S1}$.  At all baselines accessible to ground-based mm-VLBI experiments, the difference between the reconstructed and intrinsic $P^{S2}/P^{S1}$ is small in comparison to the uncertainties due to finite sample size and measurement errors.  This remains true for other power spectra ratios between polarized components (e.g., $P^Q/P^U$) on baseline lengths of interest, and at higher observation frequencies.

The improved sensitivity expected in future mm-VLBI experiments can significantly increase the accuracy with which $P^{S2}/P^{S1}$ may be measured.  More importantly, arrays like that envisioned by the ngEHT provide a much more dense sampling of the uv-plane, and therefore the ability to measure more completely the two-dimensional power spectra ratios.  The thermal uncertainties may be further reduced by aggregating nearby measurements in the uv-plane and/or exploiting assumptions regarding azimuthal symmetry; a complete discussion of these will appear elsewhere.

At higher observation frequencies, e.g., 345\,GHz, the reduced impact of scattering results in nearly exact mitigation out to baseline lengths of $\sim50\,\Gl$ and within the 25-epoch sampling variance for baseline lengths up to $\sim100\,\Gl$.  Thus it is possible to effectively mitigate interstellar scattering on baselines relevant for space-based mm-VLBI concepts that place stations in low and medium Earth orbits (2,000\,km and $<$35,000\,km, respectively), and accurately probing the magnetic field power spectrum on scales as small as $2\,\muas$ and thus a quarter of the Schwarzschild radius in Sgr A*.

\begin{acknowledgments}

We would like to thank Vedant Dhruv and many others from University of Illinois at Urbana-Champaign, who provided us with the GRMHD simulation. We would also like to thank Ramesh Narayan for his helpful comments.  This work was supported in part by Perimeter Institute for Theoretical Physics.  Research at Perimeter Institute is supported by the Government of Canada through the Department of Innovation, Science and Economic Development Canada and by the Province of Ontario through the Ministry of Economic Development, Job Creation and Trade.  A.E.B. thanks the Delaney Family for their generous financial support via the Delaney Family John A. Wheeler Chair at Perimeter Institute.  A.E.B. receives additional financial support from the Natural Sciences and Engineering Research Council of Canada through a Discovery Grant.

\end{acknowledgments}

\bibliographystyle{aasjournal_aeb}
\bibliography{main_reference.bib,EHTPapers.bib}

\clearpage

\appendix

\section{Deflection Angle vs Phase Change Caused for Different Polarization Modes}\label{appendix:DvP}

The scattering screen can be envisioned as a screen of width $L_\parallel$ comprised of many electron bubbles, which have typical size of $L_\perp$. When a photon travels travels through the scattering screen, each bubble causes a slight deflection to the photon's trajectory. After averaging the photon deflection angle over the width of the scattering screen, we should be able to derive the root-mean-squared deflection angle as a function of the width.

The argument is also valid for the phase change caused by the scattering screen, as the deflection angle and the phase change are linearly correlated. Next, we are going to consider two different scenarios, where the auto-correlation function of the electron density and the magnetic field takes different forms.

\subsection{Difference in the Deflection Angles for Different Polarization Modes}\label{a1}

We have shown in the \autoref{subsec:non-Bire} that the difference in the deflection angles and the phase changes for different polarization modes can be written as an integral over functions of the magnetic field and the electron density field, as
\begin{equation}\label{appendix:dtheta}
    \delta \theta = \abs{\theta_+ - \theta_-} = \int dz\nabla_\textbf{r} \left(XY\right),
\end{equation}
\begin{equation}\label{appendix:dphi}
    \delta \phi = \frac{\omega}{c} \int dz XY. 
\end{equation}

In this appendix, we will explore the properties of $\delta\theta$ and $\delta\phi$, given different fluctuations of the electron density fields and the magnetic fields.

We assume that $X$ and $Y$, the fluctuation electron density field and the magnetic field, are some independent Gaussian random fields. For the electron density field, it has mean value $X_0$, while for magnetic field the mean value is zero. Decomposing $X$ and $Y$ into Fourier modes with cylindrical coordinate system, we have
\begin{equation}\label{grvx}
    X\left(\r,z\right)=X_0 + \int d\q dm a_{\q,m} e^{\left[i\left(\q \cdot \r + mz\right)\right]},
\end{equation}
\begin{equation}\label{grvy}
    Y\left(\r,z\right)=\int d\q dm b_{\q,m} e^{\left[i\left(\q \cdot \r + mz\right)\right]},
\end{equation}
where $\q$ and $\r$ are the Fourier conjugate in the cylindrical plane, and $m$ and $z$ are conjugate in the axial direction.

The power spectra for the two Gaussian random fields are defined as:
\begin{equation}\label{ps}
    \langle a_{q_1,m_1} a_{q_2,m_2}^* \rangle = P\left(q_1,m_1\right)\delta\left(q_1-q_2\right)\delta\left(m_1-m_2\right)
\end{equation}
\begin{equation}\label{qs}
    \langle b_{\textbf{q}_1,m_1} b_{\textbf{q}_2,m_2}^* \rangle = Q\left(\textbf{q}_1,m_1\right)\delta\left(\textbf{q}_1-\textbf{q}_2\right)\delta\left(m_1-m_2\right).
\end{equation}

The independence of the two fields require the Fourier coefficients $a_{\q,m}$ and $b_{\q,m}$ to satisfy $\langle a_{\q_1,m_1} b_{\q_2,m_2}^* \rangle = 0$.

For simplicity, from here on in this appendix, we will use prime to denote the derivative in the radial direction, e.g. $Y' =\partial Y/ \partial r$.

Inserting the fields $X$ and $Y$ expanded in the Fourier domain back to \autoref{appendix:dtheta}, and taking the ensemble average, we have
\begin{equation}\label{eadtheta}
\begin{aligned}
    \langle \delta \theta \rangle &= \int dz \langle X'Y + \left(X-X_0\right)Y'+X_0Y'\rangle \\
    &= i \int dz \int d\q_{1,2} dm_{1,2} (\q_1 - \q_2) \\
    &\times \langle a_{\q_1,m_1} b_{\q_2,m_2}^* \rangle e^{i[(\q_1-\q_2) \cdot \r + (m_1 - m_2)z]}.
\end{aligned}
\end{equation}

A quick check can be done that because $a_{\q,m}$ and $b_{\q,m}$ to satisfy $\langle a_{\q_1,m_1} b_{\q_2,m_2}^* \rangle = 0$, the average of $\delta \theta$ is zero.

Similarly, we can insert \autoref{grvx} and \autoref{grvy} into $\langle \delta \theta^2 \rangle$, which takes the form
\begin{equation}
    \langle \delta \theta^2 \rangle = \int dz_1 dz_2 \langle \left( X_1' Y_1 + X_1 Y_1' \right)\left(X_2' Y_2 + X_2 Y_2' \right)\rangle,
\end{equation}
where the subscripts 1 and 2 of $X$ and $Y$ denote $z_1$ and $z_2$ dependence of $X$ and $Y$, and the subscripts 1, 2, 3 and 4 of $\textbf{q}$ and $m$ denote different realizations of the Gaussian random fields.

The integrand above can be divided into nine different terms, as
\begin{equation}\label{appendix:dtheta2}
\begin{aligned}
    \langle \delta \theta^2 \rangle = \int dz_1 dz_2 &\langle \left(X_1' Y_1 + \left(X_1 - X_0 \right) Y_1'+X_0 Y_1'\right)\\
    &\times\left(X_2' Y_2 + \left(X_2 -X_0\right) Y_2' + X_0 Y_2' \right)\rangle.
\end{aligned}
\end{equation}
Each nine components of the integral can be done independently using the identities \autoref{ps} and \autoref{qs}.

The first term is:
\begin{multline}
    \langle X_1' Y_1 Y_2^* X_2'^* \rangle = \int d\q_1 dm_1 d\q_2 dm_2 P\left(\q_1,m_1\right) \\
    \times Q\left(\q_2,m_2\right) \q_1^2 e^{i\left(m_1+m_2\right)\left(z_1-z_2\right)}.
\end{multline}
The second term is:
\begin{equation}
\begin{aligned}
    &\langle X_1' Y_1 Y_2'^* \left(X_2-X_0\right)^* \rangle = \\
    &\int d\q_1 dm_1 d\q_2 dm_2 P\left(\q_1,m_1\right) Q\left(\q_2,m_2\right)\\
    &\times \q_1\cdot\q_2 e^{i\left(m_1+m_2\right)\left(z_1-z_2\right)}.
\end{aligned}
\end{equation}
The third term is:
\begin{equation}
\begin{aligned}
    &\langle \left(X_1-X_0\right) Y_1' Y_2^* X_2'^* \rangle =\\
    &\int d\q_1 dm_1 d\q_2 dm_2 P\left(\q_1,m_1\right) Q\left(\q_2,m_2\right)\\
    &\times \q_1\cdot\q_2 e^{i\left(m_1+m_2\right)\left(z_1-z_2\right)}.
\end{aligned}
\end{equation}
The fourth term is:
\begin{equation}
\begin{aligned}
    &\langle \left(X_1-X_0\right) Y_1' Y_2'^* \left(X_2-X_0\right)^* \rangle =\\
    &\int d\q_1 dm_1 d\q_2 dm_2 P\left(\q_1,m_1\right) Q\left(\q_2,m_2\right)\\
    &\times \q_2^2 e^{i\left(m_1+m_2\right)\left(z_1-z_2\right)}.
\end{aligned}
\end{equation}
The fifth term is:
\begin{equation}
    \langle X_0 Y_1' Y_2'^* X_0^*\rangle = X_0^2 \int d\q dm Q\left(\q,m\right)\q^2 e^{i m \left(z_1-z_2\right)}.
\end{equation}
The other four terms are zero, because  $\langle a_{\q_1,m_1} b_{\q_2,m_2}^* \rangle = 0$.

Grouping nine terms together, we have
\begin{equation}\label{variancedtheta}
\begin{aligned}
    \langle \delta \theta^2 \rangle &= \int dz_1 dz_2 \bigg(\int d\q_1 dm_1 d\q_2 dm_2\\
    &P\left(\q_1,m_1\right) Q\left(\q_2,m_2\right)\left(\q_1+\q_2\right)^2 e^{i\left(m_1+m_2\right)\left(z_1-z_2\right)}\\
    &+ X_0^2\int d\q dm Q\left(\q,m\right) \q^2 e^{im\left(z_1-z_2\right)}\bigg).
\end{aligned}
\end{equation}

We can define the auto-correlation function of fields $X$ and $Y$ to help simplifying \autoref{variancedtheta}:
\begin{equation}
    C\left(\r,z\right) = \int d\q dm P\left(\q,m\right) e^{i\left(\q\cdot\r+mz\right)}
\end{equation}
\begin{equation}
    D\left(\r,z\right) = \int d\q dm Q\left(\q,m\right) e^{i\left(\q\cdot\r+mz\right)}.
\end{equation}

At the line of sight, \autoref{variancedtheta} can be evaluated with the auto-correlation functions $C\left(\r=0,z\right)$ and $D\left(\r=0,z\right)$:
\begin{multline}
    \langle \delta \theta^2 \rangle = -\int dz_1 dz_2 \bigg[ 
    C\left(\textbf{r},z_1-z_2\right)D\left(\textbf{r},z_1-z_2\right)\\
    + X_0^2 D\left(\textbf{r},z_1-z_2\right)\bigg]'' \Bigg|_{\textbf{r}=0}.
\end{multline}

Because the derivative is with respect of the perpendicular direction, and the integral is along the line-of-sight direction, derivative and integral can be switched:
\begin{multline}\label{vardtheta}
    \langle \delta \theta^2 \rangle = -\frac{\partial^2}{\partial r^2}\int dz_1 dz_2 \bigg[ 
    C\left(\textbf{r},z_1-z_2\right)D\left(\textbf{r},z_1-z_2\right)\\
    + X_0^2 D\left(\textbf{r},z_1-z_2\right)\bigg]
    \Bigg|_{\textbf{r}=0}.
\end{multline}

\subsection{Difference in Phase Changes for Different Polarization Modes}\label{a2}

Similar to that \autoref{appendix:dtheta} can be expanded in the Fourier space and expressed in the compact form of the auto-correlation functions, same can be done to \autoref{appendix:dphi}.

First, the ensemble average of $\delta\phi$ is zero as expected:
\begin{equation}
\begin{aligned}
    \langle \delta \phi \rangle &= \frac{\omega}{c}\int dz d\textbf{q}_1 dm_1 d\textbf{q}_2 dm_2 e^{i\left[\left(\textbf{q}_1-\textbf{q}_2\right)\textbf{r}+\left(m_1-m_2\right)z\right]} \\
    &\times \big(\langle a_{\textbf{q}_1,m_1} b_{\textbf{q}_2,m_2} \rangle + X_0\langle b_{\textbf{q}_2,m_2} \rangle 
    \big) \\
    & =0,
\end{aligned}
\end{equation}
given that $\langle a_{\q_1,m_1} b_{\q_2,m_2}^* \rangle = 0$.

Second, the variance is:
\begin{equation}
\begin{aligned}
    \langle \delta \phi^2 \rangle = \left( \frac{\omega}{c} \right)^2 \int dz_1 dz_2 
    \bigg[
    &\langle \left(X_1-X_0\right) Y_1 \left(X_2-X_0\right) Y_2 \rangle\\
    &+ \langle \left(X_1-X_0\right) Y_1 X_0 Y_2\rangle \\
    &+ \langle X_0 Y_1 \left(X_2-X_0\right) Y_2 \rangle\\
    &+ \langle X_0 Y_1 X_0 Y_2 \rangle
    \bigg].
\end{aligned}
\end{equation}
Similar to \autoref{appendix:dtheta2}, the integrand of $\langle \delta\phi^2\rangle$ can be broken into small components, whose detailed calculations are similar to the ones of $\langle\delta\theta^2\rangle$. In the end, we have
\begin{equation}
\begin{aligned}
    \langle \delta \phi^2 \rangle = \left( \frac{\omega}{c} \right)^2 &\int dz_1 dz_2 \bigg[ \int d\q_1 dm_1 d\q_2 dm_2 \\
    &P\left(\q_1,m_1\right)Q\left(\q_2,m_2\right) e^{i\left(m_1+m_2\right)\left(z_1-z_2\right)}\\
    &+ X_0^2 \int d\q_2 dm_2 Q\left(\q,m\right)e^{im\left(z_1-z_2\right)}\bigg].
\end{aligned}
\end{equation}

The equation above can be evaluated in the same way as we evaluate the variance of $\delta \theta$, with the help of the auto-correlation functions:
\begin{multline}\label{vardphi}
    \langle \delta \phi^2 \rangle = \left( \frac{\omega}{c} \right)^2 \int dz_1 dz_2 \bigg[ 
    C\left(\r,z_1-z_2\right)D\left(\r,z_1-z_2\right)\\
    + X_0^2 D\left(\r,z_1-z_2\right) \bigg]_{\r=0}.
\end{multline}
Compared to \autoref{vardtheta}, the variance of $\delta \theta$ is proportional to the second order derivative of the variance of $\delta \phi$:
\begin{equation}\label{eq:dtheta_dphi_relation}
\begin{aligned}
    \langle \delta \theta^2 \rangle
    &= \left(\frac{c}{\omega}\right)^2 \frac{\partial^2}{\partial r^2}\langle \delta \phi^2 \rangle \\
    &\approx \frac{\lambda^2}{L^2} \langle \delta \phi^2 \rangle,
\end{aligned}
\end{equation}
where $L$ is the typical correlation length within the plasma.

\subsection{Explicit Examples and Quantitative Estimates} \label{a3}

Quantitatively assessing the magnitude of the phase differences between the two polarization modes requires an explicit model for the density and magnetic field fluctuations within the scattering screen.  Because \autoref{vardtheta} and \autoref{vardphi} depend only on the correlation functions of these underlying physical quantities, specifying the statistical properties of the density and magnetic field through their correlation functions is sufficient.  Because these are not known a priori, we explore a handful of examples, beginning with a Gaussian correlation functions with a natural intrinsic scale, and culminating in the Kolmogorov models that are traditionally employed.

\subsubsection{
Gaussian Auto-correlation Functions
}

We first consider the simple case that the auto-correlation functions of the electron density, $X$, and the magnetic field, $Y$, are Gaussian.  These have a clear scale, beyond which the correlation is exponentially suppressed.
Both the Gaussian distributions have the standard deviation of $L_\perp$, which is the typical size of fluctuations in the electron density and magnetic field strength. Within each bubble, both the electron density and magnetic field fluctuations are highly correlated, while for different bubbles they are not. The auto-correlation functions are:
\begin{equation}
    C\left(\textbf{r},z\right) = \sigma_X^2 e^{-\left(r^2+z^2\right)/2L_\perp^2},
    \label{eq:GaussC}
\end{equation}
\begin{equation}
    D\left(\textbf{r},z\right) = \sigma_Y^2 e^{-\left(r^2+z^2\right)/2L_\perp^2},
    \label{eq:GaussD}
\end{equation}
where the corresponding variances $\sigma_X$ and $\sigma_Y$ are of order of one.

To estimate \autoref{vardphi}, we first make a change of variable:
\begin{equation}
    \Tilde{z_1} = z_1 - z_2
\end{equation}
\begin{equation}
    \Tilde{z_2} = z_1 + z_2, 
\end{equation}
in terms of which the variance in the phase fluctuations may be written with new upper and lower bounds:

\begin{equation} 
\begin{aligned}
    &\langle \delta \phi^2 \rangle \\ 
    &= 2\left(\frac{\omega}{c}\right)^2\int^{L_\parallel}_0 d\Tilde{z_1} \int_{\Tilde{z_1}}^{2L_\parallel-\Tilde{z_1}} d\Tilde{z_2} \left( C\left(\Tilde{z_1}\right)D\left(\Tilde{z_1}\right) + X_0^2 D\left(\Tilde{z_1}\right) \right)
\end{aligned}
\end{equation}

Inserting the Gaussian correlation functions above, this becomes,
\begin{equation}
    \begin{aligned}
    \langle \delta \phi^2 \rangle &= 4 \left( \frac{\omega}{c} \right)^2 \bigg\{ \sigma_X^2 \sigma_Y^2 \frac{L_\perp^2}{2} \bigg[\left( (-1 + e^{-L_\parallel^2/L_\perp^2}\right) \\
    & + \frac{L_\parallel}{L_\perp} \sqrt{\pi} \erf{\left(\frac{L_\parallel}{L_\perp}\right)} \bigg] + X_0^2 \sigma_Y^2 L_\perp^2 \bigg[\left( (-1 + e^{-L_\parallel^2/2L_\perp^2}\right) \\
    &+ \frac{L_\parallel}{L_\perp} \sqrt{\frac{\pi}{2}} \erf{\left(\frac{L_\parallel}{\sqrt{2}L_\perp}\right)} \bigg]\bigg\}
    \end{aligned}
\end{equation}

The scattering screen is comprised of many electron bubbles along the line-of-sight, $L_\parallel/L_\perp \gg 1$. In this limit both $\erf\left(L_\parallel/L_\perp\right)$ is approximately unity and $\exp\left(-L_\parallel^2/L_\perp^2\right)$ is approximately zero. With these simplifications, the variance of $\delta \phi$ becomes,
\begin{equation}
    \langle \delta \phi^2 \rangle = 2\left(\frac{\omega}{c}\right)^2 \sqrt{\pi}L_\parallel L_\perp \sigma_Y^2 \left(\sigma_X^2 +\sqrt{2}X_0^2\right)
\end{equation}

Similarly, the variance of the deflection angle, \autoref{vardtheta}, can be calculated given \autoref{eq:GaussC} and \autoref{eq:GaussD} in the limit that $L_\parallel/L_\perp \gg 1$.  This is facilitated by the fact that application of the second-order derivative within the screen is straightforward within the integrals.  The result is,

\begin{equation}
\begin{aligned}
    \langle \delta \theta^2 \rangle &= \int dz_1 dz_2 \bigg[ \frac{2}{L_\perp^2}C\left(\textbf{r}=0,z_1-z_2\right)D\left(\textbf{r}=0,z_1-z_2\right)\\
    &+\frac{1}{L_\perp^2}X_0^2D\left(\textbf{r}=0,z_1-z_2\right)\bigg]\\
    &=\sqrt{2\pi}\frac{L_\parallel}{L_\perp} \sigma_Y^2 \left(\sqrt{2}\sigma_X^2+X_0^2\right).
\end{aligned}
\end{equation}

These two characterizations of the degree of birefringence are related by
\begin{equation}\label{phivtheta1}
    \langle \delta \phi^2 \rangle 
    =
    4\pi^2 \frac{L_\perp^2}{\lambda^2}\langle \delta \theta^2 \rangle
    \left(
    \frac{\sqrt{2}\sigma_X^2 + X_0^2}{2\sigma_X^2+\sqrt{2}X_0^2}
    \right).
\end{equation}
where the terms in parentheses are generally of order unity.  This relationship between $\langle \delta \phi^2 \rangle$ and $\langle \delta \theta^2 \rangle$ matches the general expectation from \autoref{eq:dtheta_dphi_relation}.
\\

\subsubsection{Broken Power-law Auto-correlation Functions} \label{appendix:broken power-law}

The second case we considered is when the auto-correlation functions take the form of a broken power law, which incorporates fluctuations on multiple scales.  Above a minimum scale, $L_\perp$, the distribution of density and magnetic field fluctuations is self similar, characterized by a power-law index $\alpha$,
\begin{equation} \label{appendix:broken1}
    C\left(\textbf{r},z\right) = \frac{\sigma_X^2}{1+\left[(z^2+r^2)^{1/2}/L_\perp\right]^\alpha}
\end{equation}
\begin{equation} \label{appendix:broken2}
    D\left(\textbf{r},z\right) = 
    \frac{\sigma_Y^2}{1+\left[(z^2+r^2)^{1/2}/L_\perp\right]^\alpha}.
\end{equation}

As a concrete example, we present the computation for $\alpha=2$ prior to moving on to a Kolmogorov description; the result is qualitatively similar for any $\alpha>1$.  The variance in the phase perturbations are,
\begin{equation}
\begin{aligned}
\langle\delta\phi^2\rangle &= 2\left(\frac{\omega}{c}\right)^2\Bigg\{\sigma_X^2\sigma_Y^2\frac{1}{2}L_\parallel L_\perp \arctan\left(\frac{L_\parallel}{L_\perp}\right)\\
&+X_0^2\sigma_Y^2 \Bigg[\frac{1}{2}L_\perp^2\ln\left(\frac{L_\perp^2}{L_\perp^2+L_\parallel^2}\right)
\\
&+L_\parallel L_\perp \arctan\left(\frac{L_\parallel}{L_\perp}\right)\Bigg]\Bigg\}.
\end{aligned}
\end{equation}
Again we will assume $L_\parallel/L_\perp \gg 1$, and thus $\arctan{L_\parallel/L_\perp}\approx\pi/2$ and the logarithmic term is small relative to the linear terms.  In this limit, the phase fluctuation variance simplifies to
\begin{equation}
\begin{aligned}
\langle\delta\phi^2\rangle &= \left(\frac{\omega}{c}\right)^2\frac{\pi}{2}L_\parallel L_\perp\sigma_Y^2\left(\sigma_X^2+2X_0^2\right),
\end{aligned}
\end{equation}
which, up to factors of order unity, matches the expression found for Gaussian auto-correlation functions.

The same argument holds true also for \autoref{vardtheta}. After taking the second derivative with respect to $r$ and evaluating at $r=0$, integrating of \autoref{vardtheta} gives,
\begin{equation}
    \begin{aligned}
    \langle\delta\theta^2\rangle = \pi\frac{L_\parallel}{L_\perp}\sigma_Y^2 \left(\frac{3}{2}\sigma_X^2+X_0^2\right).
    \end{aligned}
\end{equation}
This differs only by factors of order unity from the Gaussian-correlation-function case, with
\begin{equation}\label{phivtheta2}
    \langle \delta \phi^2 \rangle 
    =
    4\pi^2 \frac{L_\perp^2}{\lambda^2}\langle \delta \theta^2 \rangle
    \left(
    \frac{\sigma_X^2 + 2X_0^2}{3\sigma_X^2+2X_0^2}
    \right).
\end{equation}
Again, this expression matches the expectation from \autoref{eq:dtheta_dphi_relation} up to factors that are generally of order unity.

This is a little different compared with that in the Gaussian case (\autoref{phivtheta1}), but as long as $\sigma_X^2$, $\sigma_Y^2$ and $X_0^2$ are of order one, \autoref{phivtheta2} also holds true for the broken power law case.

\subsubsection{Kolmogorov Turbulence-implied Auto-correlation Functions}

The autocorrelation function and the structure function describing the same field are closely correlated. A more physically inspired autocorrelation function needs to be derived from a physical model of the turbulence. Past literature has suggested the power spectrum of the electron density fluctuation (e.g. see \citet{Armstrong1995, Rickett1990}) as
\begin{equation}
    P(q)\approx C_N^2 q^{-\beta},
\end{equation}
when the scale is between the inner and outer scale, where the factor $C_N^2$ is the structural coefficient, and $\beta$ is the spectral power index, which in the Kolmogorov case, $\beta = 11/3$.

The corresponding structure function, $D_\phi$, which is the integral form of the power spectrum, is related to \autoref{StructureFunction}
\begin{equation}
D_\phi(\x)
=
\begin{cases}
\displaystyle
\left(\frac{|\x|}{x_0}\right)^2 & |\x| \ll r_{\rm in}\\
\displaystyle
\frac{2}{\alpha} \left(\frac{r_{\rm in}}{x_0}\right)^{2-\alpha}
\left(\frac{|\x|}{x_0}\right)^{\alpha}
& |\x| \gg r_{\rm in}
\end{cases}
\end{equation}
where $x_0$ is the normalization scale and $\alpha = \beta-2$ for the Kolmogorov turbulence.

Meanwhile, the autocorrelation function for any field $\phi\left(x\right)$ is
\begin{equation}
    D_\phi\left(x\right) = 2\left(\langle \phi\left(x\right)^2\rangle - C\left(x\right)\right).
\end{equation}
Given the variance of the field $\phi\left(x\right)$, $\sigma_\phi^2$, and the average value of the field, we can write down $\langle \phi\left(x\right)^2\rangle$:
\begin{equation}
    \langle \phi\left(x\right)^2\rangle = \sigma_\phi^2 + \langle \phi\left(x\right) \rangle^2.
\end{equation}

For the fluctuating fields of electron density and magnetic strength, the electron density field $X\left(\x\right)$ has the mean value $X_0$ and variance $\sigma_X^2$, and the magnetic field $Y\left(\x\right)$ has zero mean value and variance $\sigma_Y^2$. Therefore, when $|\x| \gg \r_{\rm in}$, the autocorrelation functions of the electron density and magnetic strength with Kolmogorov signature are
\begin{equation}\label{appendis:autocorrelation1}
    C\left(\textbf{x}\right) = \sigma_X^2 + X_0^2 - \frac{1}{2} \left(\frac{\abs{\textbf{x}}}{x_0}\right)^{\alpha}
\end{equation}
\begin{equation}\label{appendis:autocorrelation2}
    D\left(\textbf{x}\right) = \sigma_Y^2 - \frac{1}{2}\left(\frac{\abs{\textbf{x}}}{x_0}\right)^{\alpha},
\end{equation}

In \autoref{appendix:broken power-law}, we showed that $\langle\delta\phi^2\rangle$ and $\langle\delta\theta^2\rangle$ are linearly related by a factor of $L_\perp^2$, which we physically interpreted $L_\perp$ as the typical size of the electron bubble that the scattering screen is comprised of, if the autocorrelation functions describing the electron density and the magnetic field satisfy the form of a broken power-law. The special case we demonstrated where the power index of the broken power-law being $-2$, if expanded, falls in the regime where $|\x| \ll \r_{\rm in}$.

Similarly, for \autoref{appendis:autocorrelation1} and \autoref{appendis:autocorrelation2}, they are the first order expansion of \autoref{appendix:broken1} and \autoref{appendix:broken2} in the regime where $|\x| \gg \r_{\rm in}$:
\begin{equation} \label{appendix:broken3}
    C\left(\textbf{x}\right) = \frac{\sigma_X^2+X_0^2}{1+\left(\abs{\x}/L_\perp\right)^\alpha},
\end{equation}
\begin{equation} \label{appendix:broken4}
    D\left(\textbf{x}\right) = 
    \frac{\sigma_Y^2}{1+\left(\abs{\x}/L_\perp\right)^\alpha},
\end{equation}
where $L_\perp = 2^{1/\alpha}x_0$ can be interpreted as the typical transverse size of the electron bubble.

Similarly, inserting \autoref{appendix:broken3} and \autoref{appendix:broken4} into \autoref{vardphi} and \autoref{vardtheta} gives, respectively, 
\begin{equation}
\begin{aligned}
\langle \delta \phi^2 \rangle &= \frac{12}{5} \left( \frac{\omega}{c} \right)^2 \left( \sigma_X^2 + X_0^2 \right) \sigma_Y^2 \Bigg[ \frac{\left( \frac{L_\perp}{L_\parallel} \right)^{5/3} + 1}{\left( \frac{L_\parallel}{L_\perp} \right)^{5/3} + 1} \\
&+ \frac{L_\perp}{L_\parallel} \ln \left( 1+ \left( \frac{L_\parallel}{L_\perp}\right)^{1/3} \right) \left(2 - \frac{L_\perp}{L_\parallel}\right)\Bigg] \\
&+ 12 \left( \frac{\omega}{c} \right)^2 X_0^2 \sigma_Y^2 \Bigg[ L_\parallel L_\perp \ln \left( 1+ \left( \frac{L_\parallel}{L_\perp}\right)^{1/3} \right) \\
&- L_\perp^2 \left[\left(\frac{L_\parallel}{L_\perp}\right)^{1/3} + \ln \left( 1+ \left( \frac{L_\parallel}{L_\perp}\right)^{1/3} \right) \right] \Bigg],
\end{aligned}
\end{equation}

\begin{equation}
\begin{aligned}
\langle \delta \theta^2 \rangle &= -4 \left( \sigma_X^2 + X_0^2 \right) \sigma_Y^2 \Bigg[ 1 - \left(\frac{1}{\left( \frac{L_\parallel}{L_\perp} \right)^{5/3} + 1}\right)^2\\
&+ \frac{24}{5}\frac{L_\parallel}{L_\perp}\ln \left( 1+ \left( \frac{L_\parallel}{L_\perp}\right)^{1/3} \right) - \frac{1}{5}\left(\frac{L_\parallel}{L_\perp}\right)^{5/3} L_\perp \\
&\times \frac{8 L_\parallel \left(\frac{L_\parallel}{L_\perp}\right)^{2/3} + 13 L_\perp}{L_\parallel \left(\frac{L_\parallel}{L_\perp}\right)^{2/3} + L_\perp} \Bigg] \\
&-4 X_0^2 \sigma_Y^2 \Bigg[ 1 - \frac{1}{\left( \frac{L_\parallel}{L_\perp} \right)^{5/3} + 1} -\frac{\left(\frac{L_\parallel}{L_\perp}\right)^{2/3}}{\left(\frac{L_\parallel}{L_\perp}\right)^{2/3}+\left(\frac{L_\perp}{L_\parallel}\right)}\\
&+ 3 \frac{L_\parallel}{L_\perp} \ln \left( 1+ \left( \frac{L_\parallel}{L_\perp}\right)^{1/3} \right) \Bigg]
\end{aligned}
\end{equation}

Taking leading order in $L_\parallel$ gives
\begin{equation}
    \langle \delta \phi^2 \rangle \approx \frac{4}{5} \left(\frac{\omega}{c} \right)^2 L_\parallel L_\perp \ln\left( \frac{L_\parallel}{L_\perp} \right)\left(2\sigma_X^2 + 7X_0^2\right)\sigma_Y^2.
\end{equation}
\begin{equation}
    \langle \delta \theta^2 \rangle \approx \frac{4}{5} \frac{L_\parallel}{L_\perp} \ln\left( \frac{L_\parallel}{L_\perp} \right)\left(8\sigma_X^2 + 13X_0^2\right)\sigma_Y^2.
\end{equation}
The functional forms of $\langle \delta \theta^2 \rangle$ and $\langle \delta \phi^2 \rangle$ satisfy the same linear relation as before:
\begin{equation}
    \langle \delta \phi^2 \rangle 
    = 4\pi^2 \frac{L_\perp^2}{\lambda^2}\langle \delta \theta^2 \rangle
    \left(\frac{8\sigma_X^2 + 13X_0^2}{2\sigma_X^2+7X_0^2}\right).
\end{equation}

\section{General Inversion of Refractive Scattering}
\label{app:scatt_inverse}

\autoref{eq:scattconv} describes a linear relationship between the power spectra associated with the brightness fluctuations before and after scattering.  However, it is written in such a fashion that the observed power spectrum is a function of the properties of the scattering screen and the intrinsic power spectrum.  What is desired is an expression for $P^S_\intr(\b)$ in terms of the scattering screen and $P^S_\obs(\b)$, i.e., to invert \autoref{eq:scattconv}.  This is generally possible since $|K(\u)|>0$ and $D_\phi(\u)<\infty$ for all $\u$.  In practice, it requires some care since despite being a linear relationship, it is nonlocal due to the presence of the convolution.

Here we develop a general inversion scheme, first using a local representation of the relationship between modified correlation functions, and second making use of a perturbative expansion that exploits the limit in which the refractive scattering kernel, $\K(\u)$, is small.  \autoref{eq:Pratio_approx} makes use of the first order approximation in this peturbative solution.

\subsection{Alternate Convolution Representation} \label{sec:alt_conv_rep}

To further understand the role which $\K(\u)$ plays in the process of scattering, it is straightforward to express \autoref{eq:scattconv} in the convolution representation. In the convolution picture, we are able to acquire the solutions to the observed power spectrum $P^S_\obs(\b)$ as required. And we can further assess the impact of the scattering kernel function $\K(\u)$ on different baselines.

In the convolution representation, We define the spatial correlation functions
\begin{equation}
C^S_\obs(\x) = \int d^2\b\, e^{-i \b\cdot\x/r_F^2} P^S_\obs(\b),
\end{equation}

and
\begin{equation}
D^S_\intr(\x) = \int d^2\b\, e^{-i \b\cdot\x/r_F^2} e^{-D_\phi[\b/(1+M)]} P^S_\intr(\b).
\end{equation}
These are associated with the observed and diffractively suppressed spatial power spectra.  In terms of these, \autoref{eq:scattconv} may be written as
\begin{equation}
\begin{aligned}
C^S_\obs(\x)
&=
D^S_\intr(\x)
-
\frac{r_F^4}{4\pi^2(1+M)^6}
\frac{\partial^2 D_\phi}{\partial\x \partial\x}
:
\frac{\partial^2 D^S_\intr}{\partial\x \partial\x}\\
&=
\left(
1
-
\H:\frac{\partial^2}{\partial\x \partial\x}
\right)
D^S_\intr(\x),
\end{aligned}
\label{eq:corrrel}
\end{equation}
where 
\begin{equation}
\H \equiv \int d^2\u e^{-i\u\cdot\x/r_F^2} \K(\u) = 
\frac{r_F^4}{4\pi^2(1+M)^6}
\frac{\partial^2 D_\phi}{\partial\x \partial\x},
\label{eq:Hdef}
\end{equation}
defines a set of general coefficients across all Stokes parameters.

\subsection{General Solution}
Because $\H$ is manifestly Hermitian, \autoref{eq:corrrel} can be inverted by solving for the eigenfunctions of $(1-\H:\nabla\nabla)$, which are guaranteed to be orthogonal.  That is, solve for
\begin{equation}
\left(1 - \H:\frac{\partial^2}{\partial\x \partial\x}
\right) f_j = \lambda_j f_j
\end{equation}
from which we may obtain
\begin{equation}
D^S_\intr(\x)
= \sum_j \lambda_j^{-1} f_j(\x) \int d^2\y\, f_j(\y) C^S_\obs(\y).
\end{equation}
From this we may obtain the desired spatial power spectrum via
\begin{equation}
\begin{aligned}
P^S_\intr(\b)
&=
e^{D_\phi[\b/(1+M)]} 
\int d^2\x\, e^{2\pi i \b\cdot\x} D^S_\intr(\x)\\
&=
e^{D_\phi[\b/(1+M)]} 
\sum_j \lambda_j^{-1} \tilde{f}_j(\b) \int d^2\y\, f_j(\y) C^S_\obs(\y)\\
&=
e^{D_\phi[\b/(1+M)]} 
\sum_j \lambda_j^{-1} \tilde{f}_j(\b) \int d^2\u\, \tilde{f}_j(-\u) P^S_\obs(\u),
\end{aligned}
\end{equation}
where $\tilde{f}_j(\u)$ is the Fourier transform of the eigenfunctions.  Thus, it is not necessary to generate the correlation functions at any point.

The success of this approach depends on number of eigenmodes that must be included to approximate $D^S_\intr(\x)$ with sufficient fidelity.  

\subsection{Perturbative Solution}

Since the dependence on the eigenmodes, it may present difficulties finding general solutions without modelling the details of the scattering screen. Therefore, more practically, a pertrubative solution is preferred here.

If $\H$ is small, we may adopt a perturbative approach to inverting \autoref{eq:corrrel}.  In this approximate we expand
\begin{equation}
D^S_\intr(\x) = D^S_{\intr,0}(\x) + \epsilon D^S_{\intr,1}(\x) + \epsilon^2 D^S_{\intr,2}(\x) + \dots
\end{equation}
where $\epsilon$ is an order-counting parameter that keeps track of how many factors of $\H$ are included.  Then, we solve \autoref{eq:corrrel} at each order in $\epsilon$ assuming that $\H$ is first order:
\begin{equation}
\begin{aligned}
D^S_{\intr,0}(\x) &= C^S_\obs(\x)\\
D^S_{\intr,1}(\x) &= \H:\frac{\partial^2 D^S_{\intr,0}(\x)}{\partial\x \partial\x}  = \H:\frac{\partial^2 C^S_\obs(\x)}{\partial\x \partial\x}\\
D^S_{\intr,2}(\x) &= \H:\frac{\partial^2 D^S_{\intr,1}(\x)}{\partial\x \partial\x}\\
&= \H:\frac{\partial^2}{\partial\x \partial\x}\,
\H:\frac{\partial^2}{\partial\x \partial\x} C^S_\obs(\x)\\
&\vdots
\end{aligned}
\end{equation}
Each of these may be recast in terms of integrals over $P^S_\obs(\u)$ immediately, and upon resumming we find
\begin{equation}
\begin{aligned}
P^S_\intr(\b)
=&
e^{D_\phi[\b/(1+M)]} \bigg\{
P^S_\obs(\b)\\
&-
\int d^2\y\, \K(\y+\b):\left[\y P^S_\obs(\y)\y^T \right]\\
&+
\iint d^2\y d^2\z\, \K(\y+\b)\\
&:\left[\y \K(\z+\y):\left( \z P^S_\obs(\z)\z^T \right) \y^T \right]\\
&+ \dots
\bigg\}.
\end{aligned}
\end{equation}
Note that this may be evaluated recursively, with
\begin{equation}
\begin{aligned}
g_0(\b) &= P^S_\obs(\b)\\
g_1(\b) &= g_0(\b) - \int d^2\y\, \K(\y+\b):\left[\y g_0(\y)\y^T \right]\\
g_2(\b) &= g_1(\b) - \int d^2\y\, \K(\y+\b):\left[\y g_1(\y)\y^T \right] \\
\end{aligned}
\end{equation}
in terms of which $P^S_\intr(\b)=e^{D_\phi[\b/(1+M)]} g_\infty(\b)$.

The number of orders that must be kept depend on the size of $\H$, or equivalently $\K$, relative to the identity term.  This method has the distinct advantage of being immediately computation-ready once $Q(\u)$ is specified, without the need to solve the eigenmode problem.

\end{document}